\documentclass[aps,prl,twocolumn,notitlepage,showpacs,floatfix,superscriptaddress]{revtex4-1}
\usepackage{dcolumn}
\usepackage{tabularx}
\usepackage{bm}
\usepackage{soul}
\usepackage{amsmath,amssymb,graphicx}
\usepackage{comment}
\usepackage[colorlinks=true,citecolor=blue,urlcolor=blue,linkcolor=blue]{hyperref}
\usepackage{environ}
\NewEnviron{eqnsplit}{%
\begin{equation}
\begin{split}
  \BODY
\end{split}
\end{equation}
}

\begin{document}
\title{Electric field control of interaction between magnons and quantum spin defects}

\author{Abhishek B. Solanki}
\email{solanka@purdue.edu}
\affiliation{School of Electrical and Computer Engineering, Purdue University, West Lafayette, IN 47906, USA}
\affiliation{Birck Nanotechnology Center, Purdue university, West Lafayette, IN 47906, USA}

\author{Simeon I. Bogdanov}
\affiliation{School of Electrical and Computer Engineering, Purdue University, West Lafayette, IN 47906, USA}
\affiliation{Birck Nanotechnology Center, Purdue university, West Lafayette, IN 47906, USA}
\affiliation{Department of Electrical and Computer Engineering, University of Illinois at Urbana-Champaign, IL 60801, USA}
\affiliation{Nick Holonyak, Jr. Micro and Nanotechnology Laboratory, University of Illinois at Urbana-Champaign, IL 61801, USA}

\author{Avinash Rustagi}
\affiliation{School of Electrical and Computer Engineering, Purdue University, West Lafayette, IN 47906, USA}

\author{Neil R. Dilley}
\affiliation{Birck Nanotechnology Center, Purdue university, West Lafayette, IN 47906, USA}

\author{Tingting Shen}
\affiliation{ Department of Physics and Astronomy, Purdue University, West Lafayette, IN 47906, USA}
\affiliation{Birck Nanotechnology Center, Purdue university, West Lafayette, IN 47906, USA}

\author{Mohammad M. Rahman}
\affiliation{School of Electrical and Computer Engineering, Purdue University, West Lafayette, IN 47906, USA}

\author{Wenqi Tong}
\affiliation{School of Electrical and Computer Engineering, Purdue University, West Lafayette, IN 47906, USA}

\author{Punyashloka Debashis}
\affiliation{School of Electrical and Computer Engineering, Purdue University, West Lafayette, IN 47906, USA}
\affiliation{Birck Nanotechnology Center, Purdue university, West Lafayette, IN 47906, USA}
\affiliation{Components Research, Intel Corporation, Hillsboro, Oregon 97124, USA}

\author{Zhihong Chen}
\affiliation{School of Electrical and Computer Engineering, Purdue University, West Lafayette, IN 47906, USA}
\affiliation{Birck Nanotechnology Center, Purdue university, West Lafayette, IN 47906, USA}

\author{Joerg Appenzeller}
\affiliation{School of Electrical and Computer Engineering, Purdue University, West Lafayette, IN 47906, USA}
\affiliation{Birck Nanotechnology Center, Purdue university, West Lafayette, IN 47906, USA}

\author{Yong P. Chen}
\affiliation{ Department of Physics and Astronomy, Purdue University, West Lafayette, IN 47906, USA}
\affiliation{School of Electrical and Computer Engineering, Purdue University, West Lafayette, IN 47906, USA}
\affiliation{Birck Nanotechnology Center, Purdue university, West Lafayette, IN 47906, USA}
\affiliation{Purdue Quantum Science and Engineering Institute (PQSEI), Purdue University, West Lafayette, IN 47906, USA}
\affiliation{Institute of Physics and Astronomy and Villum Center for Hybrid Quantum Materials and Devices, Aarhus University, 8000 Aarhus-C, Denmark}
\affiliation{WPI-AIMR International Research Center for Materials Sciences, Tohoku University, Sendai 980-8577, Japan}
\affiliation{The Quantum Science Center (QSC), a National Quantum Information Science Research Center of the U.S. Department of Energy (DOE)}

\author{Vladimir M. Shalaev}
\affiliation{School of Electrical and Computer Engineering, Purdue University, West Lafayette, IN 47906, USA}
\affiliation{Birck Nanotechnology Center, Purdue university, West Lafayette, IN 47906, USA}
\affiliation{Purdue Quantum Science and Engineering Institute (PQSEI), Purdue University, West Lafayette, IN 47906, USA}
\affiliation{The Quantum Science Center (QSC), a National Quantum Information Science Research Center of the U.S. Department of Energy (DOE)}

\author{Pramey Upadhyaya}
\email{prameyup@purdue.edu}
\affiliation{School of Electrical and Computer Engineering, Purdue University, West Lafayette, IN 47906, USA}
\affiliation{Birck Nanotechnology Center, Purdue university, West Lafayette, IN 47906, USA}
\affiliation{Purdue Quantum Science and Engineering Institute (PQSEI), Purdue University, West Lafayette, IN 47906, USA}
\affiliation{The Quantum Science Center (QSC), a National Quantum Information Science Research Center of the U.S. Department of Energy (DOE)}

\date{\today}
\begin{abstract}
Hybrid systems coupling quantum spin defects (QSD) and magnons can enable unique spintronic device functionalities and probes for magnetism.  Here, we add electric field control of magnon-QSD coupling to such systems by integrating ferromagnet-ferroelectric multiferroic with nitrogen-vacancy (NV) center spins. Combining quantum relaxometry with ferromagnetic resonance measurements and analytical modeling, we reveal that the observed electric-field tuning results from ferroelectric polarization control of the magnon-generated fields at the NV. Exploiting the demonstrated control, we also propose magnon-enhanced hybrid electric field sensors with improved sensitivity. 
\end{abstract}
\maketitle


\textit{Introduction}| Hybrid platforms that combine distinct physical systems with complementary characteristics provide a unique playground to explore phenomena and device functionalities richer than their components \citep{Kurizkihybrid, lachance2019hybrid,hybridhoffman,magnonicsdavid}. Optically active quantum spin defects (QSDs), i.e. \textit{microscopic} spin impurities in insulating hosts, coupled with magnons, i.e. the elementary collective excitations of \textit{macroscopically} ordered magnetic systems, have recently emerged as one such promising hybrid spin system \citep{ andrich2017long,kikuchi2017long,trifunovic2013long,FlebusEntangleDW,trifunovic2015high,van2015nanometre,pageMR,wolfe2016spatially,zhang2020spintorque,du2017control,Mccullian, zhou2020magnon,Labanowski, Flebus2018relaxometry,finco2020AFMimaging,Avinash,Demler2019noiseMagnetometry,purser2020spinwave,backaction, Thiel}. 

The motivation for creating magnon-QSD systems is twofold.  First, magnons resonantly enhance microwave fields up to nanoscale \citep{andrich2017long}, feature long-distance nonreciprocal transport \citep{Damon} and mode confinement at reconfigurable nanoscale magnetic textures \citep{wagner,nagosa,sluka}|properties that have given birth to the burgeoning field of magnonics \citep{chumak2015magnon}. Therefore, magnons provide promising control fields for solving the challenge of on-chip coherent driving \citep{andrich2017long,kikuchi2017long,Labanowski,chunhuielectrical} and communication between QSD qubits \citep{trifunovic2013long,FlebusEntangleDW,flatte}. Second, application of well-established quantum defect magnetometry techniques to magnon-generated fields provides route to develop previously unavailable noninvasive and nanoscale probes for a broad range of magnetic phenomena \citep{andrich2017long,kikuchi2017long,trifunovic2013long,FlebusEntangleDW,trifunovic2015high,van2015nanometre,pageMR,wolfe2016spatially,zhang2020spintorque,du2017control,Mccullian, zhou2020magnon,Labanowski, Flebus2018relaxometry,finco2020AFMimaging,Avinash,Demler2019noiseMagnetometry,purser2020spinwave,backaction,Thiel}. 

Electric field control of spin system is a key resource in spintronics. Adding such electric-field control to QSD-magnon coupling would thus expand the range of phenomena and device functionalities that can be enabled by magnon-QSD hybrid systems. For example, the dependence of magnon-spin coupling on an applied electric field, when combined with the magnetometry of magnon-generated fields, provides a new scheme for sensing electric fields and phenomena via QSDs. Apart from extending the reach of QSD-based nanoscale sensors \cite{casola2018probing} to probe magnetoelectric materials, the attractive feature offered by this approach includes leveraging magnetic resonance enhancement \citep{trifunovic2015high} and the stronger magnetic field susceptibility of QSD ground state \citep{Caihybrid} to enhance the electric field sensitivity. On another front, electric fields, as opposed to magnetic fields and currents, can be confined to the scale of inter-qubit separation with minimal Joule heating \citep{Labanowski,Hideo, Bauer, Laucht}. Electric field tuning of QSD-magnon interaction could thus enable a scalable network of QSD-based circuits, where the desired spins are driven and/or entangled with their neighbors via locally tunable magnon modes.

\begin{figure*}[t]
\centering
\includegraphics[width=0.8\textwidth]{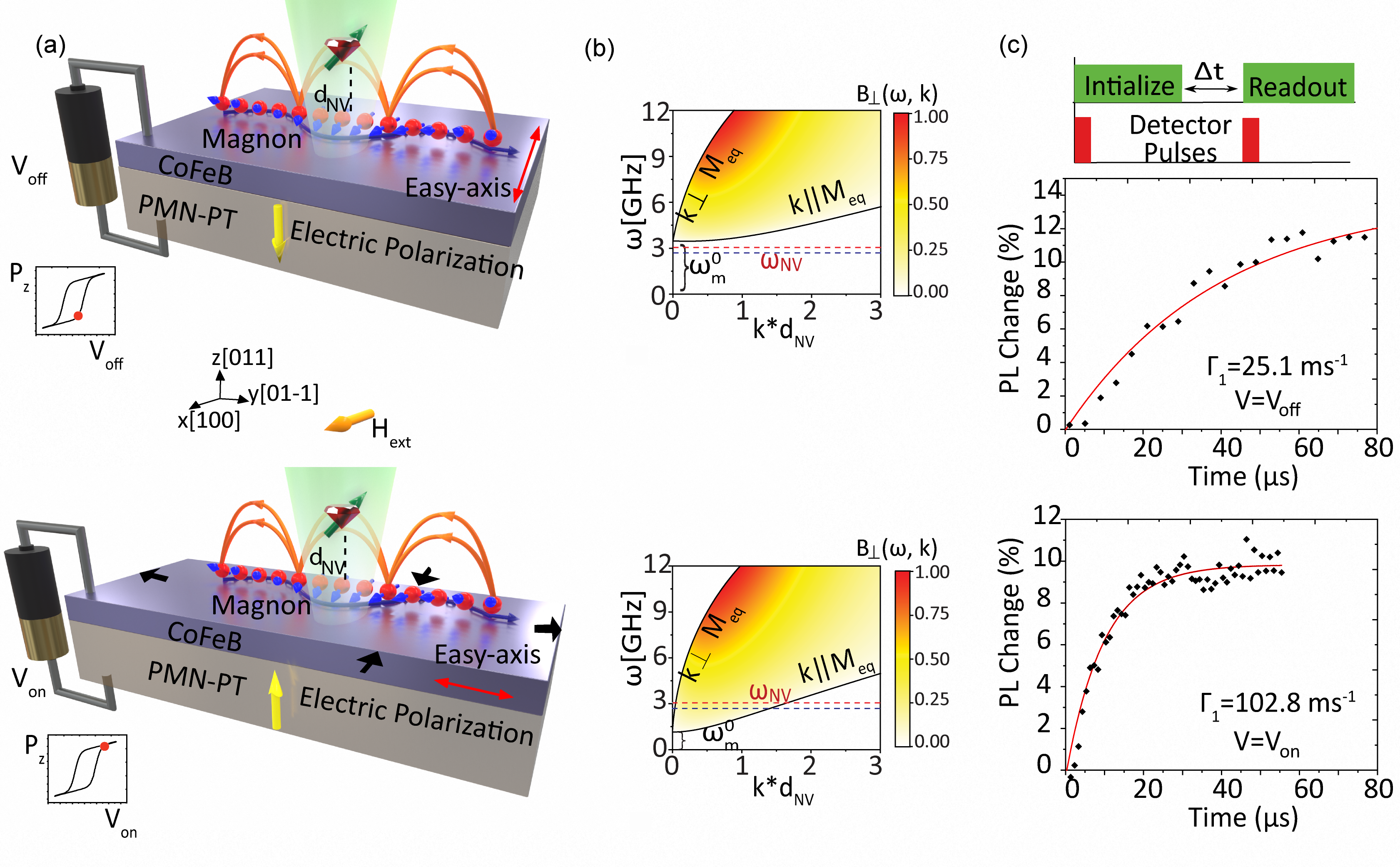}

\caption {(a) Schematic illustration of the QSD-magnon hybrid system. An external voltage $V$ controls the electrical polarization of PMN-PT (as shown in schematic P-$V$ loops) resulting in lattice strain  and change in lateral dimensions of the device structure. The direction of the magnetic anisotropy field (easy axis--red double arrows) is along $x$ for $V=V_{off}$ and along $y$ for $V=V_{on}$ in CoFeB. (b) Maps of normalized $B_{\bot}$($\emph{k}$) as a function of $\omega-\emph{k}$ for both $V_{off}$ and $V_{on}$. The black lines enveloping the colormap are the calculated magnon dispersion lines for bulk modes $(\emph{k}\parallel M)$ and surface modes $(\emph{k}\bot M)$. The dashed coloured lines represent the NV ESR lines $\omega_{NV}$.(c) Change in relaxation rates of the NV ensembles enabled by electrical tuning of QSD-magnon coupling. The schematic diagram represents the pulse sequence of the measurement scheme.}

\label{Fig:one}
\end{figure*}

In this Letter, we experimentally demonstrate electric-field control of interaction between magnons and QSDs by engineering a new hybrid system that combines a ferroelectric-ferromagnetic multiferroic \citep{NianSunPMN-PT,RameshPMN-PT,ZhangPMN-PT,tingtingPMN-PT} with QSDs. We also propose and show theoretically that the demonstrated electric field control can be used to sense electric fields with nearly 2-orders of magnitude improvement in the sensitivity when compared with sensing schemes utilizing direct coupling of ground-states of QSD with electric fields. While electric-field control of magnetism in such composite multiferroics has garnered significant attention in the classical domain \citep{NianSunPMN-PT,RameshPMN-PT}, our results highlight their utility for enabling functional quantum hybrid systems.

\textit{Central Scheme}|The device structure and the central scheme are depicted in Fig. \ref{Fig:one}(a). We disperse nanodiamonds with NV center \citep{doherty} ensembles, which act as QSDs, onto ferromagnetic (FM) (20nm) CoFeB/(300µm, 011-cut) ferroelectric (FE) PMN-PT composite multiferroic films \citep{ZhangPMN-PT,tingtingPMN-PT}. The magnons in CoFeB produce oscillating dipole magnetic fields at the NV, whose magnitude and frequencies ($\omega_m$) depends on their wavevector ($\emph{k}$) \citep{van2015nanometre}. The components of these fields that are transverse to the NV-quantization axis mediate the interaction between magnons and the NV-spins via Zeeman coupling. The central idea we demonstrate here is that by controlling the ferroelectric polarization component $P_z$ along the [011] axis in PMN-PT, magnon bands in CoFeB film can be moved with respect to the NV ESR transitions. This brings magnon-generated fields of different magnitude in resonance with the NV ESR transitions, thereby enabling electric-field control of the magnon-QSD interaction. 

We show in Fig. \ref{Fig:one}(a) the particular case of how the magnon bands respond to the flipping of $P_z$ from $-z$ to $+z$ direction in the presence of a fixed external magnetic field ($H_{ext}$) along the [100] $x$-axis. The reversal of $P_z$ is initiated at $V=V_{off}$, where the magnon bands and NV ESR transitions are off resonant, giving rise to a weak coupling. In contrast, for $V=V_{on}$ the reversal is nearly completed, and the magnon bands and NV ESR transitions are brought into the resonance, giving rise to a stronger coupling.

The above movement of magnon bands results from the coupled electric, elastic and magnetic orders in the multiferroic and can be understood as following. The (011)-cut PMN-PT features piezo-electric coefficients $d_{31}$ and $d_{32}$ of opposite signs. Consequently, to accommodate the increasing  $P_z$, PMN-PT expands along the [01-1] $y$-axis and shrinks along the [100] $x$-axis. The transfer of this anisotropic strain to CoFeB, when combined with the magnetoelastic interaction, lowers (raises) energy for the magnetization oriented along the $y\left(x\right)$ axis. This change in the electric polarization-controlled uniaxial magnetic anisotropy results in the easy-axis reorienting from $x$ to $y$ for $P_z$ changing from $-z$ to $+z$ \citep{ZhangPMN-PT,tingtingPMN-PT}. Crucially,\ both the magnon band gap $ \omega_m^0\equiv\omega_m\left(k=0\right)$ and hence the frequency range spanned by the magnon bands depend on the magnetic anisotropy \citep{kittel1948fmr}. The alignment of the direction of  $H_{ext}$ with the easy axis maximizes  $\omega_m^0$, while their orthogonal orientation minimizes $\omega_m^0$. Consequently, as the voltage is increased from $V_{off}$ to $V_{on}$ the magnon bands are pulled to lower frequencies and brought into resonance with the NV ESR transitions [see Fig. \ref{Fig:one}(b)]. 

To demonstrate and quantify the electric-field control of magnon-QSD interaction in these hybrids we perform NV- relaxometry \citep{doherty} measurements in presence of thermal magnons (see Methods in Supplementary ). At room temperature, our NV ensembles \citep{Simeon} feature an intrinsic relaxation rate of $\Gamma_0 \approx 3\ [ms]^{-1}$ due to electron-phonon interaction and paramagnetic impurities in the vicinity of NVs \citep{doherty}. The coupling to the thermal magnons acts as a source of additional magnetic noise for the NV spins. This results in an increased NV spin relaxation rate ($\Gamma_1$), which thus provides a measure of the magnon-QSD coupling. In the remainder of this Letter, we demonstrate that in NV/CoFeB/PMN-PT hybrid films magnon bands can be moved in and out of resonance with NV ESR transitions via application of the electric field. This is measured as a 400\% tuning of $\Gamma_1$ by voltage [Fig. \ref{Fig:one}(c)] in qualitative agreement with theory. We also show theoretically that this change in $\Gamma_1$ can be improved by multiple orders of magnitude by patterning and using low damping ferromagnets, which we propose to leverage for improving electric field sensing by NV.

\begin{figure}

\includegraphics[width=0.5\textwidth]{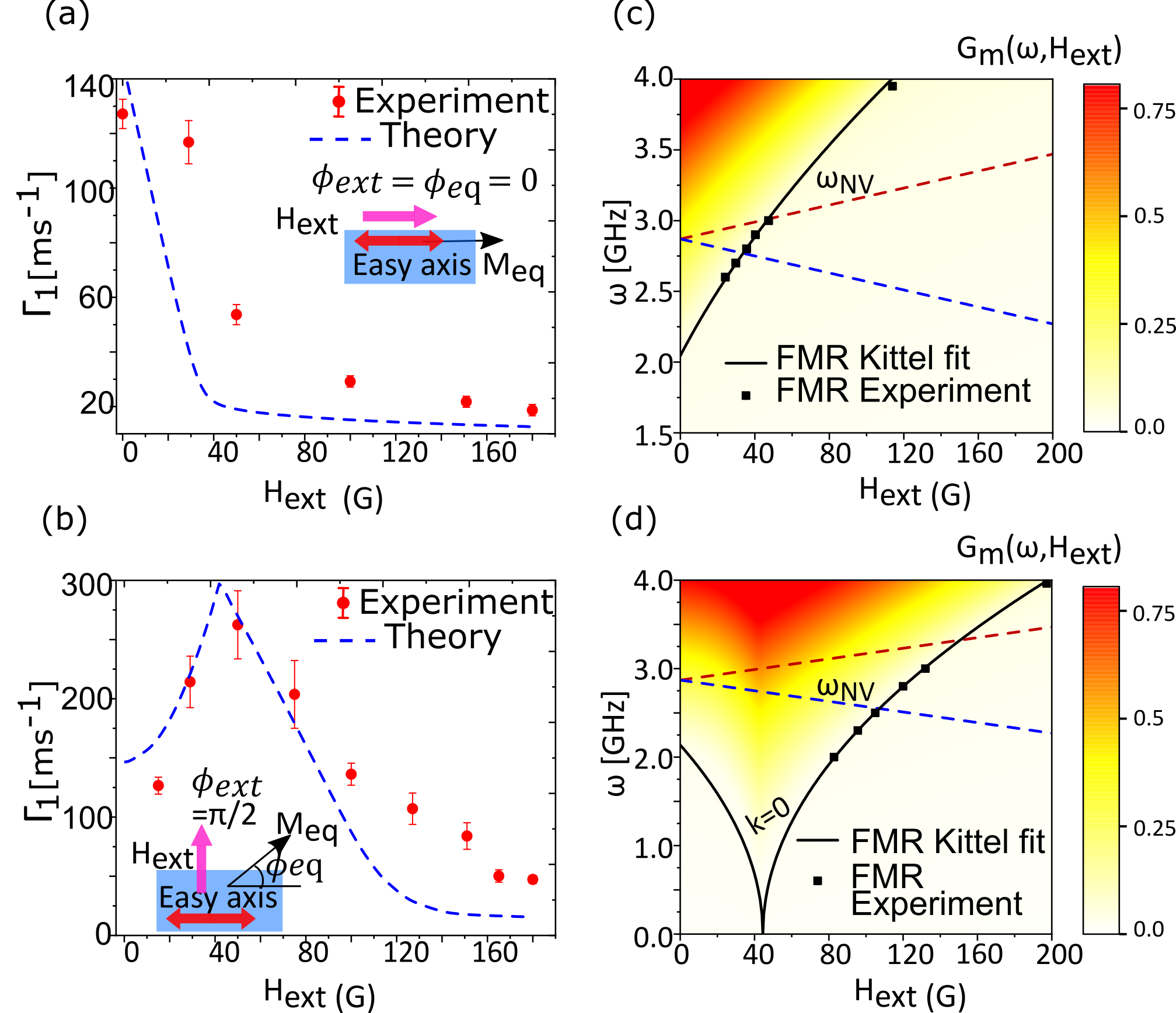}
\caption{(a), (b) Measured relaxation rate $\Gamma_1$ as a function of external magnetic field $H_{ext}$ applied parallel ($x$-axis) and orthogonal ($y$-axis) to the magnetic anisotropy field (easy-axis) respectively. The dashed lines represent theoretical fits of relaxation rates $\Gamma_1$. The inset shows the schematic variation of $H_{ext}$ w.r.t the easy-axis and the direction of equilibrium magnetization. (c), (d) Ferromagnetic resonance (FMR) frequency $({\omega}_m^0)$ as a function of external magnetic field $H_{ext}$ fitted with the Kittel formula for 20nm CoFeB film (solid lines) for $H_{ext}$ parallel ($x$-axis) and orthogonal ($y$-axis) to the anisotropy field respectively. The dashed coloured lines represent maximum spread of the NV ESR lines $\omega_{NV}$. The color map represents the calculated values of the magnetic noise spectral density  $G_m\left(\omega,H_{ext}\right)$ for an effective NV height $d_{NV}$=77nm.
}
\label{Fig:two}
\end{figure}

\textit{Anisotropic QSD-magnon coupling}.|We begin by understanding the role of a \textit{static} magnetic anisotropy field on NV-spin magnon coupling. To this end, we measure the relaxation rate $\Gamma_1$ for $H_{ext}$ applied along the [100] $x$-axis and the [01-1] $y$-axis of the (011-cut) PMN-PT [Fig. \ref{Fig:two}(a,b)]. For the field applied along the $x$-axis, $\Gamma_1$ decreases monotonically reaching the minimum value of 18.7 $\pm$ 2 $[ms]^{-1}$ for $H_{ext}=180G$. On the other hand, for the field applied along the $y$-axis, $\Gamma_1$ increases first reaching a maximum value of 262.4 $\pm$ 29 $[ms]^{-1}$ at $H_{ext}=50G$ and then decreases. The relaxation of NV-spins coupled to magnons is governed by the spectral density of magnon-generated transverse field fluctuations evaluated at the NV ESR transitions \citep{van2015nanometre}. The thermally populated modes of the magnetic film $\omega_m\left(\emph{k}\right)$ generate a magnetic noise with the spectral density given by \citep{van2015nanometre} (read Supplementary Materials: S4)	 	 

\begin{equation}
\label{eqn:one}
G_m(\omega)= \int B_\bot^2(\emph{k})D\left[\omega, \omega_m\left(\emph{k}\right)\right]n\left[\omega_m\left(\emph{k}\right)\right]Ad\emph{k}/\left(2\pi\right)^2 ,                
\end{equation}

Here, $B_{\bot}$ is the magnitude of the transverse magnetic field at the NV due to a magnon occupying the mode with wavevector $\emph{k}$ (whose relative magnitude is shown in Fig. \ref{Fig:one}b), $D=\alpha\omega_m/\pi\left[\left(\omega-\omega_m\right)^2+\alpha^2\omega_m^2\right]$ is the magnon spectral density with $\alpha$ being the Gilbert damping parameter, and $nAd\emph{k}/2\pi^2$ counts the total number of thermal magnons occupying the states in the neighbourhood of $\emph{k}$ , where $A$ is the area of the film and $n=k_BT/\hbar\omega_m$ is the Rayleigh-Jeans distribution function with $k_B$ and $T$ being the Boltzman constant and temperature, respectively. External magnetic field tunes the spectral density of magnetic noise resonant with the NV ESR transitions by controlling the magnon spectrum.

The magnon spectrum in thin films is described by the dipole-exchange spin waves \citep{kalinikos1986theory}, which can be written in the form $\omega_m\left(\emph{k}\right)=\omega_m^0+f(\emph{k})$. Here, $\omega_m^0$ (Kittel mode) is the band gap at $\emph{k}=0$ and $f(\emph{k})$ (see Supplementary Materials: S4) describes the nondegenerate branches of dispersion [c.f. magnon bands in Fig. \ref{Fig:one}(b)]. To track the location of magnon bands, in Fig. \ref{Fig:two}(c, d), we present the results of ferromagnetic resonance (FMR) experiments on our multiferroic films, which directly measure $\omega_m^0$ as a function of $H_{ext}$ along the $x$ and $y$ axis (see details in Supplementary Materials: S1). Consistent with our relaxometry measurements, $\omega_m^0$ also depends on the orientation of $H_{ext}$, lying at a higher frequency for $H_{ext}\parallel x$ when compared with that for $H_{ext}\parallel y$. This anisotropic behavior can be described by considering a uniaxial magnetic anisotropy energy of the form $\mathcal{F}_{an}=H_kM_s(m_y^2-m_x^2)/2$ \citep{tingtingPMN-PT,ZhangPMN-PT,NianSunPMN-PT} in the magnetic film. Here, $M_s$ is the saturation magnetization, $m_x$ and $m_y$ are the $x$ and $y$ components of the unit vector oriented along the magnetization, and $H_k$ parameterizes the strength of the uniaxial anisotropy field. When $H_k>0 (H_k<0)$, the easy axis is oriented along the $x$ ($y$) axis. In such films, $\omega_m^0$ is governed by the Kittel formula \citep{kittel1948fmr}:  

\begin{equation}
\label{eqn:two}
\omega_m^0=\gamma\sqrt{H_1H_2}, 
\end{equation}

with $H_1=H_{ext}\cos{\left(\phi_{ext}-\phi_{eq}\right)}+2H_k\cos{2\phi_{eq}}$ and $H_2=H_{ext}\cos{\left(\phi_{ext}-\phi_{eq}\right)}+H_k\cos{2\phi_{eq}}+4\pi M_s$. Here, $\phi_{ext}$ and $\phi_{eq}$ are the azimuthal angles of the external magnetic field and the equilibrium magnetization, respectively [see inset Fig. \ref{Fig:two}(a,b)]. The Kittel formula fits are shown in Fig. \ref{Fig:two}(c,d), which gives $H_k=20G$ with the easy-axis oriented along the $x$-axis. The central result highlighted by these fits is that $\omega_m^0$ and thus the magnon bands shift monotonically to higher frequencies for $H_{ext}$ parallel to the $x$-axis. In contrast, for $H_{ext}$ along the $y$-axis, $\omega_m^0$ and the band frequencies are pulled down first, before rising to higher values. 

Equipped with the magnon spectrum, we plot the normalized field dependent noise spectral density $G_m\left(\omega, H_{ext}\right)$, alongwith the maximum spread of NV ensemble ESR frequencies $\omega_{max}^+=2.87+\gamma H_{ext}[GHz]$ and $\omega_{min}^-=2.87-\gamma H_{ext}[GHz]$ in Figs. \ref{Fig:two}(c) and \ref{Fig:two}(d) (see Supplementary Materials: S4). For $H_{ext}\parallel x$, magnons generating higher $B_{\bot}$ are pushed away from the frequency range probed by the NV ensemble due to the monotonic shift of magnon bands to higher frequency with $H_{ext}$ [see Fig. \ref{Fig:one}(b)]. As a result, for $H_{ext}\parallel x, G_m\left(\omega,\ H_{ext}\right)$ within $\omega_{max}^+$ and $\omega_{min}^-$ decreases monotonically, consistent with the relaxation rate’s decrease with the field. Whereas, for $H_{ext}\parallel y$,  since $\omega_m^0$ is first pushed to lower frequencies for $H_{ext}\ <50G$  before moving to higher frequencies for $H_{ext}> 50G$, magnons with stronger $B_{\bot}$ are brought into (out of) resonance with NV ensemble ESR transitions for  $H_{ext}<50G ( H_{ext}>50G)$. Consequently, $G_m\left(\omega,\ H_{ext}\right)$ within $\omega_{max}^+$ and $\omega_{min}^-$ and the relaxation rates increase (decrease) with $H_{ext}$ for $H_{ext}<50G ( H_{ext}>50G)$. 

Theoretical best fits of relaxation rate (see Supplementary Materials: S4) are also shown in Fig. \ref{Fig:two}(a, b) exhibiting qualitative agreement with the experimental results. Quantitative differences may arise from neglecting finite mode ellipticity for magnons \citep{van2015nanometre,Avinash}, two-magnon scattering-induced NV-relaxation \citep{Mccullian}, and/or effect of local inhomogeneities (arising, for example, from local strain) on the magnon spectrum.

\begin{figure}[hbtp]
\centering
\includegraphics[width=0.5\textwidth]{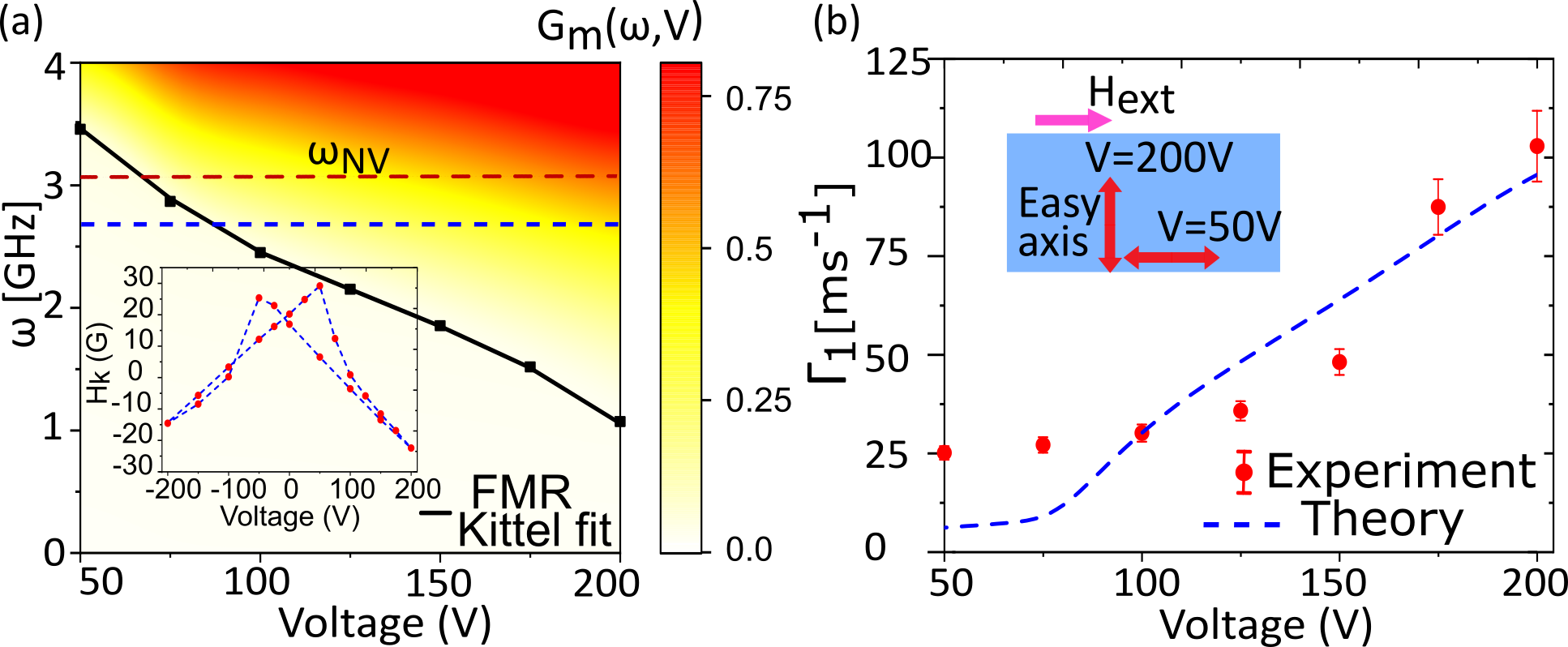}
\caption{(a) FMR frequency $(\omega_m^0)$ (data in black lines) as a function of applied voltage extracted from the experimental results for a fixed external magnetic field $H_{ext} = 57G$ along $x$-axis. The color map represents the calculated values of the magnetic noise spectral density $G_m(\omega,V)$ for an effective NV height $d_{NV}$=77nm. The dashed coloured lines represent maximum spread of the NV ESR lines $\omega_{NV}$. The inset shows the detailed measurements of magnetic anisotropy field as a function of applied voltage. We begin by polling the ferromagnet to (-200V) and then change the voltage in steps towards +200V. Following this, the measurement is performed by changing the voltage in reverse direction from 200V to -200V. (b) Measured relaxation rates $\Gamma_1$ as a function of applied voltage for a fixed $H_{ext}=57G$ along $x$-axis. The inset shows a schematic illustration of magnetic anisotropy field for the two different voltages for a fixed $H_{ext}$. The dashed line represents theoretical fit of relaxation rates $\Gamma_1$.}

\label{Fig:three}
\end{figure}

\textit{Electric field control}.|The multiferroic hybrid offers the attractive property of moving magnon spectrum with electric field by \textit{dynamically} tuning the magnetic anisotropy. In Fig. \ref{Fig:three}(a), we show $\omega_m^0(V)$ for $H_{ext}=57G$ along with $H_k\left(V\right)$ (see inset) extracted by applying Kittel formula Eq. \ref{eqn:two} on the measured  $\omega_m^0\left(V\right)$ (see Supplementary Materials: S1). The $H_k\left(V\right)$ inherits the characteristic butterfly-shaped curve of anisotropic strain as a function of applied voltage, which is the hallmark of coupled ferroelectric, elastic and magnetic order tuning the anisotropy in FM/FE multiferroics \citep{ZhangPMN-PT,tingtingPMN-PT,NianSunPMN-PT}. Particularly, as the voltage is increased from 50V to 200V, $H_k$ decreases monotonically from $H_k=30G$ at $V$=50V to $H_k=-22G$ at $V$ = 200V. Notably, this maximal change in $H_k$ corresponds to the flipping of the easy axis from $x$ to $y$ consistent with flipping of $P_z$ in PMN-PT (see Supplementary Materials: S2). Correspondingly, we observe the largest $\omega_m^0=3.45GHz$ when the easy-axis is aligned with the direction of $H_{ext}$ for $V$=50V. Conversely, $\omega_m^0$ decreases to the minimum value $1.07 GHz$ when the easy-axis is aligned orthogonal to the direction of $H_{ext}$ at $V$=200V. In the following, we focus on this polarization flipping-induced changes in NV-spin and magnon coupling.

We obtain the normalized field dependent magnetic noise spectral density $G_m\left(\omega,\ V\right)$ shown in Figs. \ref{Fig:three}(a) by substituting the electric field-dependent magnon spectrum from FMR experiments into Eq. \ref{eqn:one} (see Supplementary Materials: S4). On the same figure we also depict the NV ESR frequencies, which lie within the dashed horizontal lines $\omega_{max}^+\left(H_{ext}=57G\right)$ and $\omega_{min}^-\left(H_{ext}=57G\right)$ and remain unaffected by applied electric field. For $V$=50 V, i.e. $P_z$ points downward, the $\omega_m^0= 3.45GHz$ is the largest for our hybrids and lies above the NV transition frequency [Fig. \ref{Fig:three}(a)] [as shown in Fig. \ref{Fig:one}(b), where we referred to $V$=50V as $V_{off}$]. Consequently, magnons are off-resonant with NVs and are expected to couple weakly with them, which is reflected in the low value of calculated $G_m\left(\omega,\ V\right)$  within $\omega_{max}^+\ and\ \omega_{min}^-\ $ band at $ V$=50V. As the voltage is increased, the polarization of PMN-PT is reoriented, $H_k$ and $\omega_m^0$ decrease and magnons with stronger dipole fields are brought into the resonance with the NV ESR transitions [see Fig. \ref{Fig:one}]. As a result, the magnons interact strongly with the NV spins. This increased coupling is reflected in $G_m\left(\omega,\ V\right)$  increasing monotonically and reaching the maximum value within the frequency range spanned by $\omega_{max}^+$ and $\omega_{min}^-$ for $V$ = 200V; we thus refer to $V$ = 200V as $V_{on}$. 

The experimentally measured spin relaxation rate indeed increases monotonically as the voltage varies from $V_{off}$ to $V_{on}$ [see Fig. \ref{Fig:three}(b)], showing good agreement with the above qualitative picture.  Namely, the lowest relaxation rate of 25.1 $\pm$ 2 $[ms]^{-1}$ is observed for $V=V_{off}$ which increases monotonically to 102.8 $\pm$ 9 $[ms]^{-1}$ at $V=V_{on}$ .The corresponding optical spin contrast curves are shown in Fig. \ref{Fig:one}(c). Finally, we also plot the results of theoretical $\Gamma_1$ fits for the NV ensembles, with the nanodiamond orientation found in the previous section (see Supplementary Materials: S4). The theoretical fits are in reasonable quantitative agreement with the experiments. In addition to the reasons pointed out for Fig. \ref{Fig:two}(a, b), the quantitative difference between experiment and theory in this case may arise from additional inhomogeneous magnon-scattering potentials introduced by ferroelectric domains due to domain-nucleation mechanism of ferroelectric switching \citep{Li2017PMN-PT} which can be addressed by using NV-scanning geometry \citep{finco2020AFMimaging}. Furthermore, we also performed control experiments with hybrid structures of NV-ensembles coupled with PMN-PT substrate without CoFeB film (Supplementary Materials: S3) and observed no measurable voltage modulation of the NV relaxation time.

\textit{Discussion}.|The observed shift in NV’s relaxation rates, $\Gamma_1$, can be exploited for sensing E-fields (hereafter, ‘magnon-sensing’). To quantify the performance of magnon-sensing, we compare its expected single-spin DC E-field sensitivity to the sensing schemes based on direct coupling of NV’s ESR levels \citep{dolde2011} to the DC electric field (hereafter, ‘direct-sensing’). The sensitivity of magnon-sensing and direct-sensing schemes scales as $S\propto1/\eta\sqrt{T_\chi}$ \citep{Degan}; (see Supplementary Materials: S5). Here, $\eta$ is the transduction parameter, which for the magnon-sensing (direct-sensing) is given by $\eta_m=\partial\Gamma_1/\partial E$ ($\eta_d$=Stark shift \citep{dolde2011}), and $T_{\chi}$ is the time scale before which unwanted signals dephases or relaxes the NV-spin. For direct-sensing, $\eta_d\sim28Hz/\left(V/cm\right)$\citep{dolde2011} and $T_{\chi}$~ms is the NV’s decoherence time, which yields photon shot noise-limited single-spin DC field sensitivities $S_d\sim890\left(V/cm\right)/\sqrt{Hz}$ \citep{dolde2011}. For our experiments, $\eta_m \sim15Hz/\left(V/cm\right)$ and $T_{\chi}=1/\Gamma_1~10\mu s$ from the high voltage bias region of Fig. \ref{Fig:three}(b). This suggests that the sensitivity of the magnon-sensing scheme $S_m$, is an order of magnitude poorer than $S_d$ for the parameter regime we study here. However, since $\eta_m$ is not limited by the weak Stark shifts, $S_m$ can be improved by several orders of magnitude by properly designing the ferromagnet as we highlight next.

The transduction parameter for magnon-sensing can be written as $\eta_m=\nu\beta$, where $\beta\equiv\partial\omega_m/\partial E$ parameterizes the shift in magnon modes due to E-field, and the dimensionless factor $\nu\equiv\partial\Gamma_1/\partial\omega_m$ represents the change in the NV-relaxation corresponding to the shift of magnon modes. $\beta$ is governed by the piezoelectric and magneto-elastic properties of the CoFeB/PMN-PT stack, which can be read from Fig. \ref{Fig:three}(a) to be $\sim0.5MHz/\left(V/cm\right)$, about four orders larger than $\eta_d$. Therefore, small $(\sim15\ Hz/(V/cm))$ values of $\eta_m$ in our present hybrids arises from $\nu\ll1$. Physically,  $\nu\ll1$ means that a broad band of magnons must be moved before NV and magnons become off-resonant [c.f. Fig. \ref{Fig:one}(b)], resulting in slow variation of $\Gamma_1$ with frequency.

In nanoscale magnets (instead of films) the magnon modes are discretized due to confinement \citep{Guo}. Consequently, a much smaller shift in the magnon frequency ($\sim\alpha\omega_m$) would make the magnon mode and the NV ESR off-resonant, hence enhancing $\nu$. Additionally, choosing a smaller $\alpha$ would further enhance $\nu$. Informed by these heuristics, we calculate $S_m$ for low damping CoFeB \citep{Leelowdamping,schoenlowdamping} nanomagnet/PMN-PT hybrids, yielding minimum sensitivity $S_{m}\sim1\ (V/cm)/\sqrt{Hz}$ (see Supplementary Materials: S5). Other knobs to improve $S_m$ include choosing magnets with larger magneto-elastic coefficients ($\beta$), such as Terfanol-D \citep{NianSunPMN-PT,terfenol} and/or lower Gilbert damping ($\alpha$) such as YIG \citep{epiYIG} and VTCN \citep{VTCNE}. Furthermore, our proposed scheme can also leverage various other mechanisms for electric field control of magnetism \citep{Hideo,Bauer} such as voltage control of interfacial magnetic anisotropy and magneto-ionic effect to enhance $\beta$. 

In summary, leveraging coupled ferroelectric, lattice, magnetic and spin degrees of freedom, we demonstrated dynamic tuning of magnon and QSD interaction by electric fields. As an example application of the demonstarted control, we also proposed magnetic resonance-enhanced electric field sensors. Our relaxometry data suggests that the QSD relaxation can be used as a noninvasive probe of ferroelectric state via magnetoelectric coupling. Future research extending the present study to the scanning NV geometry \citep{Thiel} could thus extend the advantages of NV-center probing of condensed matter \citep{casola2018probing} to multiferroics and a broad range of magnetoelectric phenomena studied in spintronics. Beyond sensing, an array of QSD/nano-FM/FE \citep{chenMTJ,BiswasPMN-PT} hybrids could provide a novel approach to build scalable QSD-based quantum circuits. In such circuits, selective driving of QSD-qubits by a weak global microwave drive, and tunable transfer of information between such qubits, can potentially be activated electrically by tuning the magnon frequency of nano-FMs in proximity to the desired qubits.

\textit{Acknowledgements} | This work was supported in part by NSF Award 1838513, NSF Award 1944635, NSF 2015025-ECCS and the U.S. Department of Energy, Office of Science through the Quantum Science Center (QSC), a National Quantum Information Science Research Center.



\bibliography{main_v2}

\end{document}


\rmfamily

\title{ Supplementary - Electric field control of interaction between magnons and quantum spin defects }

\maketitle
\date{\today}

\section{Materials and Methods}

\subsection{Sample Preparation}
The multiferroic heterostructure stack was prepared by depositing amorphous CoFeB film on one side of poled (011) cut PMN-PT substrate (from ctscorps) in a magnetron sputtering system with a base pressure of $3\times10^{-8}$ Torr. Following this a 5nm Ta capping was deposited in the same chamber. The bottom electrode was formed by depositing a 10nm Ta and 100nm Au film in e-beam evaporation system on the other side of the PMN-PT substrate. The nanodiamonds used in the experiment were purchased from Adamas Technologies Inc, in form of an aqueous suspension of 0.1\% w/v. As per the supplier specifications, each nanodiamond consists of 400 NVs. The average size of the nanodiamonds scanned by Scanning Electron Microscopy is 76nm with a standard deviation of 21nm over 22 samples \citep{Simeon}. The solution is ultra-sonicated for an hour to avoid the formation of clumps of nanodiamonds and dispersed over the sample followed by drying in a vacuum desiccator box. 

\subsection{Relaxation rate measurement}
The relaxation rate was measured using a home built confocal microscopy setup. The sample is illuminated by a 532nm laser focused on a ~1$\mu$m spot size via a 100x objective (Nikon TU Plan Fluor EPI 0.9NA) which is also used to collect the photoluminescence (PL). The 532 nm laser is modulated using an accousto-optical modulator (AOM, Gooch and Housego) to produce 10$\mu$s pulses. The PL signal is sent through a 50$\mu$m pinhole and filtered using a dichroic mirror and a long pass 550 nm filter to remove the scattered 532 nm photons and measured with a single photon detector connected with a high speed counter (Measurement Computing USB-CTR04). The AOM and the counter are gated with a delay generator (Quantum Composers 9528). External magnetic field ($H_{ext}$) is applied through a permanent magnet. The laser power at the sample in the reported measurements is 25$\mu$W.  
 A suitable nanodiamond with sufficient PL is chosen to ensure that the measurement is not limited by photon shot noise. The relaxation rate is retrieved from the experimental data collected by following the well-known pulse scheme depicted in Fig. 1(c) of the main text. The NV ensemble is initialized in spin $m_s$=0  state by optical pumping (532nm), followed by a variable delay time $\Delta$t. At the end of the delay time, the NV spin state is determined by measuring spin-dependent photoluminescence. The optical pulse sequence is repeated $10^5$ times for each data point to improve the signal to noise ratio. The NV ensemble reaches thermal equilibrium state after enough time delay resulting in a saturation in measured PL rate. The contrast between PL rate after initializing in $m_s$=0 and PL rate after delay time $\Delta$t is fit to an exponential equation $C\left(\Delta t\right)=C_0\left(1-\exp{\left(-\Gamma_1\Delta t\right)}\right)$ to extract the spin relaxation rate.

 \newpage
 
 \section{Supplementary Text}

In Supplementary Fig. S1. we show a Photoluminescence map with nanodiamonds dispersed over a $10\times10 \mu m^2$ area. We also show an example of a measured Relaxation rate data for nanodiamonds with NV Ensemble dispersed over a glass coverslip in Supplementary Fig. S2.

\subsection{\underline{S1:Ferromagnetic Resonance Measurements (FMR)}}
The frequency of the lowest energy, non-propagating, uniform precession magnon mode is called the ferromagnetic resonance (FMR) frequency ($\omega_m^0$). Ferromagnetic resonance measurement is a popular technique for measuring magnetic anisotropy field, gyromagnetic ratio and gilbert damping in ferromagnetic materials. We used Physical Properties Measurement System (PPMS) by Quantum Design Inc. to perform measurements. External magnetic field is applied in the plane of the magnet. The microwave magnetic field is delivered to the sample by a coplanar waveguide structure. 
The functional dependence of FMR frequency on magnetic anisotropy field ($H_k$) and external magnetic field ($H_{ext}$) is explained in Eq. (2) in the main text. We first measure the FMR frequency ($\omega_m^0$) vs $H_{ext}$ for $V$=0. Microwave energy is absorbed when the frequency of applied microwave magnetic field matches the $\omega_m^0$ for a fixed applied magnetic field (or vice-versa). As an example, Supplementary Fig. S3 shows an absorption curve for a fixed microwave frequency f=5GHz as the external magnetic field is varied. The Gilbert damping factor $\alpha$ is extracted from the linewidth of the absorption curve which goes as $\alpha\omega_m^0$. The measured $\omega_m^0$ vs $H_{ext}$ is fitted with the Kittel formula (Eq. (2) main text) to extract saturation magnetization ($M_s$) and magnetic anisotropy field ($H_k$). 

Next, we measure FMR curves for different voltages. We begin by polling the ferroelectric substrate to -200V. FMR curves are measured for different voltages changed in steps from -200V to +200V and then reverse from +200V to -200V. Each dataset is fitted with the Kittel Formula to extract magnetic anisotropy field $H_k (V)$. The resultant fits are also used to extract $\omega_m^0$  ($H_{ext}$=57 G $\hat{x}$) for each voltage point because the NV relaxation measurements are done for  $H_{ext}$=57 G $\hat{x}$. Complete FMR measurements for Voltage $V$=50V to $V$=200V fitted with Kittel Formula are shown Supplementary Fig. S4,S5,S6,S7.

\subsection{\underline{S2:Electric Polarization of the single crystal PMN-PT}}

Single crystal PMN-PT (011 cut) was employed in this work to take advantage of its large anisotropic piezoelectric coefficients $d_{31}$ $\sim$~-3100pC/N along [100] direction and $d_{32}$ $\sim$ 1400pC/N along the [01-1] direction when the DC voltage is applied along the [011] direction \citep{tingtingPMN-PT}. Measured ferroelectric polarization as a function of DC voltage as shown in Supplementary Fig. S8 demonstrates a typical hystersis behavior. The ferroelectric polarization was interpreted by measuring surface charge as a function of applied DC Voltage with Radiant RT66C. Change in ferroelectric polarization is also accompanied by a large strain which is reflected in a change in magnetic anisotropy field shown in Fig. 3a of the main text.

\subsection{\underline{S3:Control Measurements}}

We performed control measurements to support our interpretation of results presented in the main text. We dispersed nanodiamonds containing NV ensembles on a PMN-PT substrate without the CoFeB magnetic film. We performed measurements of NV relaxation rate ($\Gamma_1$) for different values of magnetic field applied in the plane of the magnet ($H_{ext}$) and Voltage (V) applied across the substrate as shown in Supplementary Fig. S9 and Fig. S10 respectively. 

First, we note that the typical values of relaxation rate ($\Gamma_1$) on PMN-PT substrate is much smaller than the values observed with CoFeB as shown in the main text. This result highlights that magnon-NV coupling dominates the relaxation rate of the NV Ensembles. We do not observe significant dependence of the relaxation rate $\Gamma_1$ on $H_{ext}$ in comparison to the case when the NVs are allowed to interact with magnons in the CoFeB film presented in Fig. 2 of the main text. However, we observe a significant dependence of saturation contrast $(C(t\rightarrow\infty)=C_0)$ 
on the magnitude of the external magnetic field. However, we use two fitting parameters $C_0$ and $\Gamma_1$ to fit our experimental results to the equation stated before $C(t)=C_{0}(1-\exp(-\Gamma_{1}t))$  which ensures that the extracted values of $\Gamma_1$ is reliable and doesn't depend on magnetic field if the CoFeB film is absent. 

We do not observe any significant dependence of the relaxation rate $\Gamma_1$ on the external voltage ($V$) in absence of the CoFeB film. These measurements strongly support our interpretation that the NV Ensembles interact with magnons to produce the dependence on magnetic and electric field demonstrated in Fig. 2 and Fig. 3 of the main text.

\subsection{\underline{S4:Analytical calculation of the magnon spectrum and NVE Relaxation rates}}
\subsubsection{Free Energy}
The free energy describing the ferromagnetic thin film of CoFeB is governed by Zeeman, uniaxial anisotropy, exchange, and demagnetization terms
\begin{equation}
\mathcal{F} = -M_s \vec{H}_{ext} \cdot \vec{m} +{H}_{k} \dfrac{M_s}{2} \left( m_y^2-m_x^2\right) + A_{ex} [(\partial_x \vec{m})^2 + (\partial_y \vec{m})^2] - \dfrac{M_s}{2} \vec{m} \cdot \vec{H}_D,
\label{free}
\end{equation}
where $\vec{H}_{ext}$ is the external field applied in the $x$-$y$ plane, $M_s$ is the saturation magnetization, $A_{ex}$ is the exchange energy constant, ${H}_{k}$ is the uniaxial magnetic anisotropy such that when ${H}_{k} > 0$ the easy axis is along the $x$-axis and when ${H}_{k} < 0$ the easy axis is along $y$-axis, and $\vec{H}_D$ is the demagnetization field given in Fourier domain by
\begin{equation}
\begin{split}
\vec{H}_D = -\hat{G}(\vec{k}) \vec{m}(\vec{k}) &= -\left(\begin{array}{ccc}
G_{xx}(\vec{k}) & G_{xy}(\vec{k}) & 0 \\ 
G_{yx}(\vec{k}) & G_{yy}(\vec{k}) & 0 \\ 
0 & 0 & G_{zz}(\vec{k})  
\end{array}  \right) \left(\begin{array}{c}
m_x(\vec{k}) \\ 
m_y(\vec{k}) \\ 
m_z(\vec{k})
\end{array}\right) \\
&= -H_d \left(\begin{array}{ccc}
\cos^2\phi_k \, g_k & \sin\phi_k \cos\phi_k \, g_k & 0 \\ 
\sin\phi_k \cos\phi_k \, g_k & \sin^2\phi_k \, g_k & 0 \\ 
0 & 0 & 1-g_k 
\end{array}  \right) \left(\begin{array}{c}
m_x(\vec{k}) \\ 
m_y(\vec{k}) \\ 
m_z(\vec{k})
\end{array}\right), \\
\end{split}
\end{equation}
where $H_d = 4\pi M_s$, $L_z$ is the film thickness and $g_k = 1 - (1-\exp(-k L_z))/(kL_z)$. Depending on the magnitude and direction ($\phi_{ext}$) of the external field relative to the anisotropy field, the equilibrium magnetization can have orientation in the $x$-$y$ plane characterized by the azimuthal angle $\phi_\mrm{eq}$ as will be addressed in the next section.

\subsubsection{Magnetization Equilibrium}
\paragraph{Case 1: No Voltage + External Field along in-plane Easy axis}

Here $H_{k} > 0$ i.e. easy axis is along the $x$-axis. In this situation, the equilibrium magnetization is also along the $x$-axis i.e. $\phi_{eq}=0$.

\paragraph{Case 2: No Voltage + External Field along in-plane Hard axis}

Here ${H}_{k} >0$ i.e. easy axis is along the $x$-axis and external field is applied along the $y$-axis $\phi_{ext} = \pi/2$. In this situation, the equilibrium magnetization is given by a conditional expression,
\[ \phi_{eq} = \begin{cases} 
      \pi/2 & H_{ext} > 2 H_k \\
      \sin^{-1}(H_{ext}/2H_k) &  H_{ext} \leq 2 \vert H_k \vert \\
      \end{cases}
.\] \\
\paragraph{Case 3: With Voltage + External Field along $x$-axis}

Here $H_{k}$ is modulated by voltage with positive (negative) implies $x$- ($y$-)axis is the easy axis and external field is applied along the $x$-axis $\phi_{ext} = 0$. In this situation, the equilibrium magnetization is given by a conditional expression,
\[ \phi_{eq} = \begin{cases} 
      0 & H_k \geq 0 \\
      0 & H_k <0 \, \text{and} \, H_{ext} \geq 2 \vert H_k \vert \\
      \cos^{-1}\left(H_{ext}/2\vert H_k \vert  \right) & H_k <0 \, \text{and} \, H_{ext} \leq 2 \vert H_k \vert 
   \end{cases}
.\] \\

\subsubsection{Spinwave Dispersion} 
The magnetization dynamics of the magnetic moments in the ferromagnetic thin film in absence of dissipation is described by the Landau-Lifshitz equation
\begin{equation}
\label{LL}
\dfrac{d}{dt} \vec{m} = - \gamma \, \vec{m} \times \vec{H} ,
\end{equation}
where $\vec{m}$ is the magnetization unit vector, $\gamma>0$ is the Gyromagnetic ratio, and $\vec{H}=-M_s^{-1} \partial \mathcal{F}/\partial\vec{m}$ is the effective magnetic field experienced by the magnetization. To solve for the spin-wave dispersion using the linearized Landau-Lifshitz equation, we need to rotate the coordinate system such that the $x$-axis aligns with the equilibrium magnetization (we label this rotated frame variables by prime in the superscript). This can be achieved by rotating about the $z$-axis by an angle $\phi_{eq}$ captured by the matrix
\begin{equation}
R_z(\phi_{eq}) = \left( \begin{array}{ccc}
\cos\phi_{eq} & \sin\phi_{eq} & 0 \\ 
-\sin\phi_{eq} & \cos\phi_{eq} & 0 \\ 
0 & 0 & 1
\end{array} \right),
\end{equation}
such that
\begin{equation}
\vec{m}^\prime = R_z(\phi_{eq}) \vec{m}  \quad \text{and} \quad \vec{m} = R^T_z(\phi_{eq}) \vec{m}^\prime .
\end{equation}
In the rotated frame 
\begin{equation}
\vec{m}^\prime_{eq} = \left( \begin{array}{c}
1 \\ 
0  \\ 
0
\end{array} \right) \quad \text{and} \quad \delta\vec{m}^\prime = \left( \begin{array}{c}
0 \\
\delta m_y^\prime \\ 
\delta m_z^\prime  
\end{array} \right) .
\end{equation}
Therefore,
\begin{equation}
\delta\vec{m} = R^T_z(\phi_{eq}) \delta\vec{m}^\prime = \left( \begin{array}{c}
-\sin\phi_\mrm{eq} \delta m_y^\prime \\
\cos\phi_\mrm{eq} \delta m_y^\prime \\ 
\delta m_z^\prime  
\end{array} \right).
\end{equation}
The effective magnetic field experienced by the magnetic moments in Fourier space is
\begin{equation}
\vec{H} = H_{ext} [\cos\phi_{ext} \hat{x} + \sin\phi_{ext} \hat{y}] + H_k [m_x \hat{x} - m_y \hat{y}] - H_{ex} k^2 \vec{m} + \vec{H}_D ,
\end{equation}
where $H_{ex}$ is the effective magnetic field due to exchange interaction term in the free energy expression in Eq. ~\ref{free}.

The linearized Landau-Lifshitz equation can be written in the rotated frame as
\begin{equation}
\dfrac{d}{dt} \delta\vec{m}^\prime = - \gamma \, \vec{m}_{eq}^\prime \times \delta \vec{H}^\prime - \gamma \, \delta\vec{m}^\prime \times \vec{H}_{eq}^\prime ,
\end{equation}
which for the components of transverse magnetization deviations is
\begin{equation}
\begin{split}
\dfrac{d}{dt} \delta {m}_y^\prime = \gamma \, \delta {H}_z^\prime - \gamma \, \delta {m}_z^\prime {H}_{eq,x}^\prime , \qquad \dfrac{d}{dt} \delta {m}_z^\prime = - \gamma \, \delta {H}_y^\prime + \gamma \, \delta {m}_y^\prime {H}_{eq,x}^\prime .
\end{split}
\end{equation}
The equilibrium and the deviation field in the rotated frame can be evaluated using the rotation matrix
\begin{equation}
\vec{H}_{eq}^\prime = \left( \begin{array}{c}
H_{ext} \cos(\phi_{ext} - \phi_\mrm{eq}) + H_k \cos2\phi_\mrm{eq} \\
H_{ext} \sin(\phi_{ext} - \phi_\mrm{eq}) - H_k \sin2\phi_\mrm{eq} \\ 
0 
\end{array} \right) ,\qquad \delta\vec{H}^\prime = \left( \begin{array}{c}
\cos\phi_\mrm{eq} \, \delta H_x + \sin\phi_\mrm{eq} \, \delta H_y \\
-\sin\phi_\mrm{eq} \, \delta H_x +\cos\phi_\mrm{eq} \, \delta H_y \\ 
\delta H_z
\end{array} \right), 
\end{equation}
where the required components of the deviation field in the linearized LLG are evaluated to be
\begin{equation}
\begin{split}
\delta H_y^\prime &= -\left[ H_k \cos2\phi_\mrm{eq} + H_{ex} k^2 + H_d g_k \left( \cos^2\phi_k \sin^2\phi_\mrm{eq} + \sin^2\phi_k \cos^2\phi_\mrm{eq} - \dfrac{\sin2\phi_k}{2} \sin2\phi_\mrm{eq} \right) \right] \delta m_y^\prime ,\\
\delta H_z^\prime &= -\left[ H_{ex} k^2 + H_d (1-g_k) \right] \delta m_z^\prime .
\end{split}
\end{equation}
The LLG equations in frequency domain is thus,
\begin{equation}
\begin{split}
-i\omega \delta {m}_y^\prime = - \omega_1 \delta {m}_z^\prime \qquad -i\omega \delta {m}_z^\prime = \omega_2 \delta {m}_y^\prime ,\\
\end{split}
\end{equation}
where
\begin{equation}
\begin{split}
\omega_1 &= \gamma \left[ H_{ext} \cos(\phi_{ext} - \phi_\mrm{eq}) + H_k \cos2\phi_\mrm{eq} + H_{ex} k^2 + H_d (1-g_k) \right] ,\\
\omega_2 &= \gamma \left[ H_{ext} \cos(\phi_{ext} - \phi_\mrm{eq}) + 2 H_k \cos2\phi_\mrm{eq} + H_{ex} k^2 + H_d g_k \left( \cos^2\phi_k \sin^2\phi_\mrm{eq} + \sin^2\phi_k \cos^2\phi_\mrm{eq} - \dfrac{\sin2\phi_k}{2} \sin2\phi_\mrm{eq} \right) \right]  .\\
\end{split}
\label{dispersion}
\end{equation}
Thus, the spinwave dispersion is given by $\omega_m (\vec{k}) = \sqrt{\omega_1 \omega_2}$. This expression represents the complete magnon spectrum for a given magnitude and direction of $H_{ext}$. In the main text, we write this spectrum as $\omega_m (\vec{k}) = \omega_m^0+ {\sqrt{\omega_1 \omega_2} - \omega_m^0} \equiv  \omega_m^0 + f(k)$, where $\omega_m^0 \equiv \omega_m (\vec{k}=0)$ is the bandgap at $\vec{k}=0$. 

\subsubsection{Magnon Noise and NV Transition Rates}
Within the linearized Landau-Lifshitz equation framework, the variation of magnetic moments in real space can be expressed as
\begin{equation}
\vec{m} = \vec{m}_\mrm{eq} + \sum_{\vec{k}} \delta \vec{m} \, e^{i \vec{k}\cdot\vec{r}}.
\end{equation}
Using the dipolar tensor, we can find the dipolar field at the NV (located at a height $d_{NV}$ above the magnetic thin film) in the lab frame (i.e. coordinate system in the schematic)
\begin{equation}
\vec{H}_{dip} (\vec{k}) \sim M_s \mathcal{D}(\vec{k}) \delta \vec{m} .
\end{equation}
where the dipolar tensor \citep{Avinash}
\begin{equation}
\begin{split}
\mathcal{D}(\vec{k}) &= A_k \left( \begin{array}{ccc}
\cos^2 \phi_k & \sin2\phi_k/2 & -i\, \cos\phi_k \\ 
\sin2\phi_k/2 & \sin^2\phi_k & -i\, \sin\phi_k \\ 
-i\, \cos\phi_k  & -i\, \sin\phi_k  & -1
\end{array} \right)
\end{split}
\end{equation}
and $A_k= e^{-k d_{NV}} \left[1-e^{-kL}\right]$ which in the thin film limit \citep{guslienko2011magnetostatic} $kL \ll 1$ implies $A_k \approx kL \,e^{-k d_{NV}-kL} $. \\

However, to compute the relaxation dynamics of NV, we need to find field components transverse to the NV axis and relate them to the linearized deviation of the magnetization. For this, we need to again use the rotation matrices,
\begin{equation}
\begin{split}
\vec{H}^{NV} &\sim M_s R_{NV}(\theta_{NV},\phi_{NV}) \mathcal{D}(\vec{k}) R^T_z(\phi_{eq}) \delta\vec{m}^\prime = M_s \mathcal{D}^{eff} (\vec{k})\delta\vec{m}^\prime ,
\end{split}
\end{equation}
where $\mathcal{D}^{eff} (\vec{k}) = R_{NV}(\theta_{NV},\phi_{NV}) \mathcal{D}(\vec{k}) R^T_z(\phi_{eq}) $ and 
\begin{equation}
R_{NV}(\theta_{NV},\phi_{NV})=R_{y}(\theta_{NV})R_{z}(\phi_{NV}) = \left( \begin{array}{ccc}
\cos\theta_{NV} \cos\phi_{NV} & \cos\theta_{NV} \sin\phi_{NV} & -\sin\theta_{NV} \\ 
-\sin\phi_{NV} & \cos\phi_{NV} & 0 \\ 
\sin\theta_{NV} \cos\phi_{NV} & \sin\theta_{NV} \sin\phi_{NV} & \cos\theta_{NV}
\end{array} \right) .
\end{equation}

Under the influence of the fluctuating magnetic field noise, the initialized quantum spin of the NV relaxes. The Hamiltonian for the NV (spin-1) in presence of magnetic field is given by
\begin{equation}
\mathcal{H} = D \, S_z^2 + \gamma \, \vec{S} \cdot \vec{H}^{NV} ,
\end{equation}
where $\vec{S}$ is the spin-1 operator, and the zero field splitting of the $m_s = \pm 1$ state with respect to the $m_s=0$ state is denoted as $D = (2\pi) 2.87$ GHz. A static magnetic field  splits the $\pm 1$ states by Zeeman coupling via the component of static field along the NV axis. This defines an effective set of two-level systems interacting with the magnetic field noise. The effective two-level Hamiltonian for an NV with the ESR transition  $\omega$ can then be written as,
\begin{equation}
\mathcal{H(\omega)} = \frac{\hbar \omega}{2} \sigma_z+ \dfrac{\gamma}{\sqrt{2}} [H^{NV}_x \sigma_x + H^{NV}_y \sigma_y],
\end{equation}
where $\vec{\sigma}$ are the Pauli matrices. Using Fermi’s golden rule, we can evaluate the transition rate $\Gamma_1$ governed by the magnetic field noise-induced transitions between the two levels described by the above Hamiltonian as (we have assumed without loss of generality that the NV is located at $\vec{r}$=0)
\begin{equation}
\begin{split}
\Gamma_{1}(\omega) &= \dfrac{\gamma^2}{2} \int dt \, e^{i\omega t} [\langle H^{NV}_x (t) H^{NV}_x (0) \rangle + \langle H^{NV}_y (t) H^{NV}_y (0) \rangle ] \\
&= \dfrac{\gamma^2}{2} \int dt \, e^{i\omega t} \sum_{\vec{k}} [\langle H^{NV}_x (\vec{k},t) H^{NV}_x (\vec{k},0) \rangle + \langle H^{NV}_y (\vec{k},t) H^{NV}_y (\vec{k},0) \rangle ] .\\
\end{split}
\label{def_spec}
\end{equation}
Here, $G_m(\omega) \equiv \int dt \, e^{i\omega t} [\langle H^{NV}_x (t) H^{NV}_x (0) \rangle + \langle H^{NV}_y (t) H^{NV}_y (0) \rangle ] $ is the spectral density of the transverse magnetic field noise produced at the NV due to thermally excited magnons.  

In line with the previous works \citep{du2017control,van2015nanometre,purser2020spinwave}, the NV transition rates can be expressed as
\begin{equation}
\Gamma_{1}(\omega) = \sum_{\vec{k}} D(\omega , \omega_m(\vec{k})) \mathcal{F}_m(\vec{k}) = A \int \dfrac{d{\vec{k}}}{(2\pi)^2} D(\omega , \omega_m(\vec{k})) \mathcal{F}_m (\vec{k}) .
\end{equation}
where we have converted the sum over wavevector to an integral with $A$ being the planar area of the thin film, and  $D(\omega , \omega_m(\vec{k}))$ is the magnon spectral density given by
\begin{equation}
D(\omega , \omega_m(\vec{k})) = \dfrac{1}{\pi} \dfrac{\alpha \omega_m}{(\omega - \omega_m)^2 + \alpha^2 \omega_m^2} \xRightarrow[\alpha \rightarrow 0]{} \delta(\omega - \omega_m)
\end{equation}
and
\begin{equation}
\mathcal{F}_m (\vec{k}) = \gamma^2 M_s^2 \delta m_y^{\prime \, 2} [ \vert \mathcal{D}_{xy}^{eff} (\vec{k}) \vert^2 + \vert \mathcal{D}_{yy}^{eff} (\vec{k}) \vert^2].
\end{equation}
The magnetization deviations has an in-plane component and a transverse out-of-plane component. Given, the large demagnetization energy cost for out-of-plane deviations, we have only considered the in-plane magnetization deviation $\delta m_y^\prime$. Thus, the magnetization deviation $\delta m_y^\prime$ is related to the thermally excited magnons by the following,
\begin{equation}
M_x^{\prime \, 2} = M_s^2 - M_s^2 [\delta m_y^{\prime \, 2}+\delta m_z^{\prime \, 2}] \approx M_s^2 - M_s^2 \, \delta m_y^{\prime \, 2} \Rightarrow M_x^{\prime} = M_s - \dfrac{M_s}{2} \delta m_y^{\prime \, 2} .
\end{equation}

This deviation of magnetization from equilibrium can be related to the change in magnetization from the thermal magnons,
\begin{equation}
M_x^{\prime} = M_s - \dfrac{n[\omega_m(\vec{k})] \gamma \hbar}{V} , 
\end{equation}
where the number of magnons $n[\omega_m(\vec{k})] = k_B T / \hbar\omega_m(\vec{k})$. As a result 
\begin{equation}
\delta m_y^{\prime \, 2} = \dfrac{2 n[\omega_m(\vec{k})] \gamma \hbar}{M_s V} .
\end{equation}
Hence, the transition rates are given by
\begin{equation}
\Gamma_{1}(\omega) = C_s \,\dfrac{2 \hbar \gamma^2 M_s}{V }  \int [ \vert \mathcal{D}_{xy}^{eff} (\vec{k}) \vert^2 + \vert \mathcal{D}_{yy}^{eff} (\vec{k}) \vert^2] D(\omega , \omega_m(\vec{k})) n[\omega_m] \dfrac{A d{\vec{k}}}{(2\pi)^2} .
\end{equation}
where, as typically done in literature \citep{du2017control,van2015nanometre,purser2020spinwave}, we have added a scaling parameter $C_s$ to account for any inherent approximations that we made in terms of ignoring the out-of-plane component of magnetization deviation. We note that the ratio of area and volume appearing ensures the result being independent of the lateral dimensions of the magnetic thin film. We can re-arrange this rate expression in the form 
\begin{equation}
\Gamma_{1}(\omega) = C_s \frac{\gamma^2}{2} \, \int B_\perp^2(\vec{k})D(\omega , \omega_m(\vec{k}))n[\omega_m]  \dfrac{A d{\vec{k}}}{(2\pi)^2} ,
\label{rate}
\end{equation}
where
\begin{equation}
B_\perp^2(\vec{k}) = \dfrac{4 \hbar M_s}{V } [ \vert \mathcal{D}_{xy}^{eff} (\vec{k}) \vert^2 + \vert \mathcal{D}_{yy}^{eff} (\vec{k}) \vert^2] 
\label{B}
\end{equation}
can be identified as the square of the amplitude of the dipolar field generated at NV by a magnon with wavevector $\vec{k}$. Comparing Eq.~(\ref{rate}) with Eq.~(\ref{def_spec}), we get:
\begin{equation}
G_m(\omega) = \, C_s\int B_\perp^2(\vec{k})D(\omega , \omega_m(\vec{k}))n[\omega_m]  \dfrac{A d{\vec{k}}}{(2\pi)^2} .
\label{spec}
\end{equation}
Eqs.~(\ref{rate}), (\ref{B}), (\ref{spec}) are the main equations that are used to generate theoretically calculate normalized magnetic field maps, normalized spectral density maps and fits of relaxation rate to experiments presented in the main text. In particular, since we are using an ensemble of NVs, we average these over the four possible NV axis orientations as defined by the host diamond lattice. Using the tetrahedral crystal structure of the diamond lattice, we can find all the possible NV orientations given one (as they are constrained by the angle of 109.5$^\circ$ between the bonds of the diamond lattice). The three unknowns - i.e. scale factor $C_s$ and the NV orientations characterized by angles $\theta_{NV}$ and $\phi_{NV}$- are determined via fits of total relaxation rate $\Gamma_1=\Gamma_1(\omega=\omega_+)+\Gamma_1(\omega=\omega_-)$ to the experimentally measured rates.

\subsection{\underline{S5:Electric field sensing: nanomagnet geometry}}

In this section, we show that by optimizing the magnet geometry and use of lower damping magnets, the sensitivity of the proposed magnon sensing scheme can be enhanced by nearly 2-3 orders of magnitude. The proposed setup consists of a nanomagnet (CoFeB) placed on top of multiferroic material (PMN-PT) (Fig. S11). Here, instead of the magnetic film geometry studied experimentally, we focus on the case of a nanomagnet geometry. The nanomagnets are expected to offer a sharper change in NV relaxation rate as a function of the electric field. This is because nanoscale confinement results in a set of discrete magnetic resonance modes being coupled to the NV (as opposed to a band of magnons coupled to the NV for the case of the film geometry).

For simplicity, we consider the case where the equilibrium orientation of magnet, NV and externally applied magnetic field are all pointing in the same direction. Within the macrospin approximation, the free energy in this case can be written by neglecting $A_{ex}$ in Eq.~\ref{free} as-
\begin{equation}
\label{free_energy}
    \mathcal{F} = -M_s \vec{m}.{\vec{H}_{ext}} +{H}_{k} \dfrac{M_s}{2}(m_y^2 - m_x^2) + \frac{M_s}{2}H_{D} m_z^2,
\end{equation}
where we have additionally focused on the geometry where the in-plane dimensions of the magnet are much larger than the out-of-plane dimensions. Consequently only out-of-plane dipolar contribuitions $\sim m_z^2$ are kept in the free energy of the nanomagnet. The magnon resonance frequency considering these interactions can thus be calculated by solving the linearized Landau-Lifshitz equation, and given by the Kittel formula obtained by substituting $k=0$ in Eq. ~\ref{dispersion}- 
\begin{equation}
\label{FMR_mode}
    \omega_{m}^0 = \gamma \sqrt{(H_{ext} + H_k) (H_{ext} + H_k + H_{D}')}
\end{equation}
where  $H_{D}' = 4\pi M_s - \dfrac{H_k}{2}$. For the dimensions that we are interested in, higher order $k\neq 0$ modes of the nanomagnet are further away from the QSD transition frequencies \citep{Guo} which allows us to focus only on the magnet FMR mode. Thus, in order to compute the NV relaxation rates, we need to calculate the magnetic field created by the FMR mode at the location of the NV. Focusing only on the $m_s=0$ and $m_s=-1$ levels as an example and using the Fermi’s golden rule, the NV transition rate $\Gamma_1(\omega)= \frac{\gamma^2}{2} G_m(\omega)$ in this case can be written as \citep{Avinash}: 
\begin{equation}
\begin{split}
\Gamma_{1}(\omega) &= \dfrac{\gamma^2}{2} \int dt \, e^{i\omega t} \langle H^{NV}_+ (t) H^{NV}_- (0) \rangle
\label{nanogamma}
\end{split}
\end{equation}
where $H_\pm=H_x\pm H_y$. We note that, different from the case of films, we have now kept magnetic field noise generated at the NV center by both out-of-plane and in-plane magnetization deviations to get rid of the unknown scaling factor $C_s$ \citep{Avinash}.

In order to estimate the above relaxation rates, we need to calculate the field-field correlators. For this, we first calculate the magnetization dynamics by solving the stochastic LLG equation numerically with the free energy as expressed in Eq. ~\ref{FMR_mode}. A Runge-Kutta solver with a timestep of 2 ps was used to solve the magnetization dynamics. To emulate the effect of ensemble average, the results were averaged over 500 runs. The dipolar field at the location of NV was calculated by following the procedure described in \citep{trifunovic2015high}. 

 Fig. S12 shows the relaxation rate of NV for the proposed setup obtained by substituting the numerically calculated field field correlators into Eq.~\ref{nanogamma}. Here, we have used a 750 $ nm \times$ 750 $ nm \times$ 10 $ nm$ magnet with a NV height, $d_{NV} =$90 $nm$. In addition, we have used a Gilbert damping parameter of $\alpha =0.001$, as observed for epitaxial CoFeB films \citep{Leelowdamping}. As expected, in contrast to magnet films of our experiments, the relaxation rate changes sharply near resonance. This results from the combined action of mode discretization (due to nanoscale mode confinement) and lower linewidth of these modes (due to low damping). Consequently, the maximum transduction parameter $\eta_m=\nu \beta \sim 1240$ (see Fig. S12) becomes nearly \textit{two orders of magnitude} larger than our experimentally observed value. We note that for this estimate we have assumed the same value of $\beta=\partial \omega/\partial E$ as observed in our experiments. Indeed, reduction of damping without negatively affecting magnetoelastic properties (which governs $\beta$)
 have been demonstrated by annealing the films \citep{zighem}. Assuming similar $\beta$ can be achieved for even lower damping insulating magnets, such as YIG with $\alpha \sim 10^{-4}-10^{-5}$, a further 1-2 order of enhancement in $\eta_m$ is possible. This makes strain-tunable and low-damping nanomagnets attractive candidates for E-field sensing applications.  

Next, following the procedure described in \citep{Degan} we estimate the minimum E-field sensitivity ($S_m$) for the QSD/multiferroic hybrids. Under the current protocol (Fig. S11), the NV is initialized to $m_{s}=0$ state and then left to evolve under the noise generated by the magnet. Therefore, the population of $m_{s}=0$ state can be written as- 
\begin{equation}
    p_0(t) = \frac{1}{2} + \frac{1}{2}\exp[-2\Gamma_1(E) t].
\end{equation}
Here, $\Gamma_1(E)$ is the NV relaxation rate to be modulated by E-field via its interaction with the nanomagnet. For small DC E-field, $\Gamma_1(E)$ can be approximated as-
\begin{equation}
    \Gamma_1(E) = \Gamma_{1}^0 + \frac{\partial\Gamma_1}{\partial E} E.
\end{equation}
Thus, the absolute value of difference in population between the presence and absence of the electric field signal becomes $|\delta p| = e^{-2\Gamma_1^0t}Etd\Gamma_1/dE $. The noise $\sigma_p$ associated with the entire readout process can be attributed to the quantum projection noise as well as the classical readout noise. Thus, we can write the signal to noise ratio (SNR)-
\begin{equation}
    SNR = \frac{|\delta p|}{\sigma_p} = 2e^{-2\Gamma_1^0t}EtC\sqrt{\frac{T}{t+t_m}}\frac{d\Gamma_1}{dE}.
\end{equation}
Here we have used \citep{Degan}: $\sigma_p = \sqrt{t+t_m}/2C\sqrt{T}$, where $C$ is a dimensionless constant quantifying readout efficiency, $t_m$ denotes the time required to initialize and read the qubit state and $T$ is the total available measurement time. By setting SNR $=$ 1 and choosing a total integration time of 1 s, we can write the minimum measurable E-field per second, i.e. sensitivity-
\begin{equation}
    S_m = \frac{\exp[2\Gamma_1^0t] \sqrt{t+t_m}}{2Ct \eta_m }.
\end{equation}
Here, we have used $\eta_m = \partial \Gamma_1/\partial E = \nu \beta$ as the transduction parameter for magnon-based sensing. For estimating optimum $S_m$, we choose the regime $t_m \lesssim t$ and $t \approx 1/\Gamma_1^0$ \citep{Degan}. In this regime, we get $S_m \sim \sqrt{\Gamma_1^0}/2C\eta_m$. Biasing NV at the frequency where $\eta_m =\partial \Gamma_1/\partial E$ peaks  (see Fig. S12), using a typical  $C\sim 0.3$ for NV read via traditional photoluminescence-based techniques \citep{hopper}, we get a minimum sensitivity of the order of $\sim$ 1 V/cm/$\sqrt{Hz}$ for a single NV. The detail optimization of this sensitivity as a function of nanomagnet size, NV placement, external magnetic field, and the chosen protocol will be presented elsewhere.

\bibliography{supplementary}

\newpage

\begin{figure}

    \centering

        \includegraphics[width=0.4\textwidth]{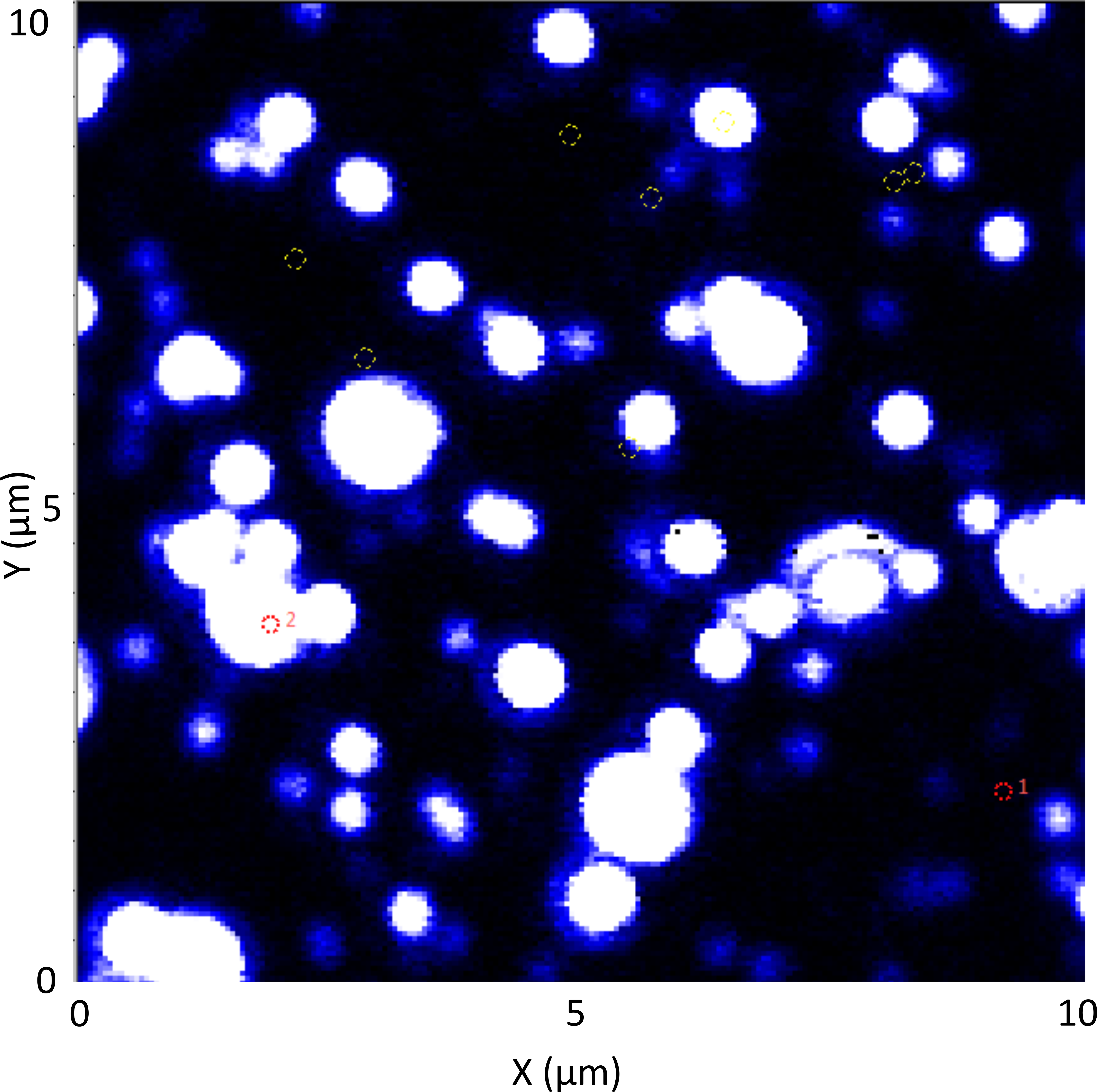}
        \caption*{\textbf{Fig. S1.} Photoluminescence map of a 10$\times$10 $\mu m^2$ area. The bright spots are nanodiamonds}
        \label{fig:galaxy}
\end{figure}
   
\newpage    
\newpage
\begin{figure}
        \includegraphics[width=0.6\textwidth]{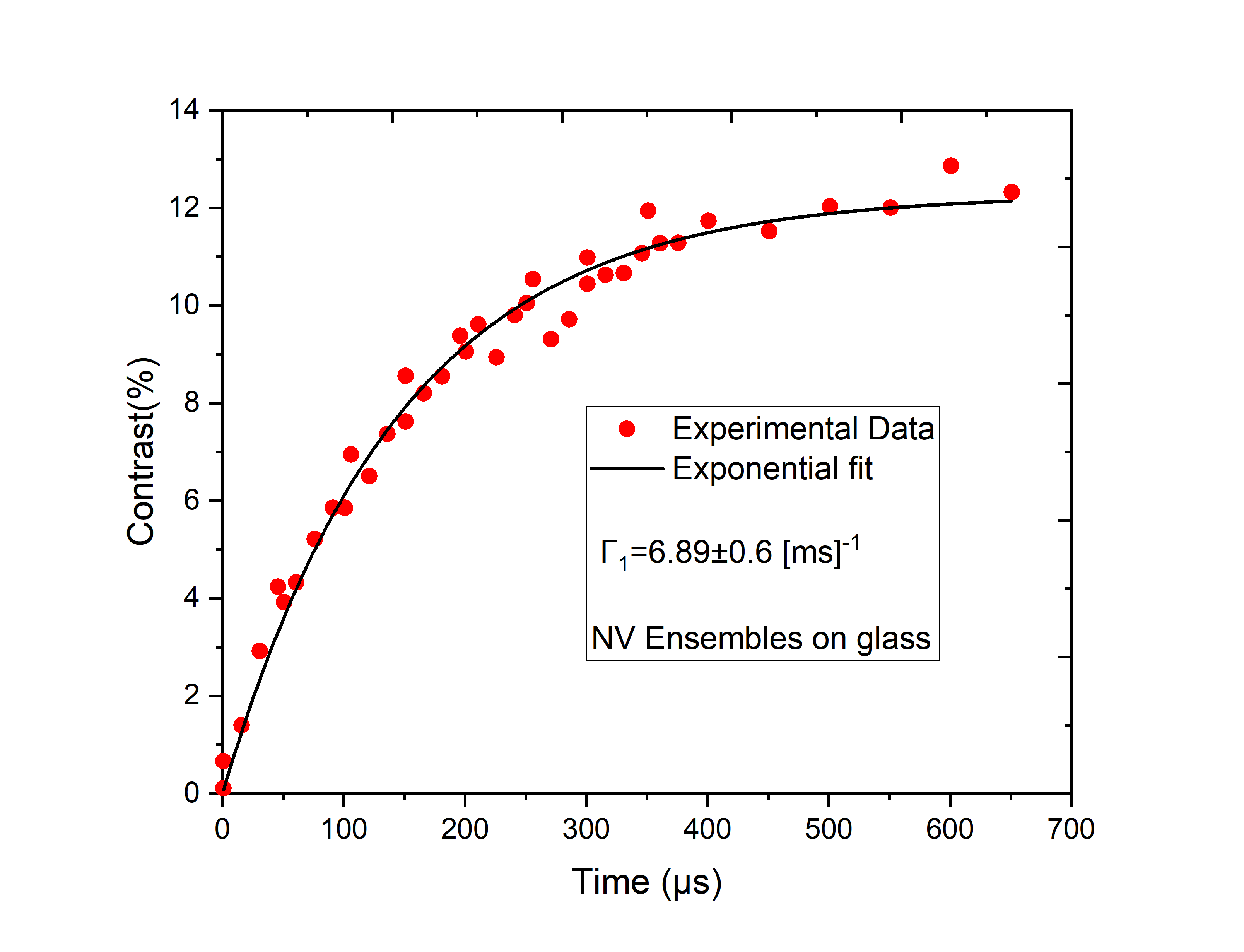}
        \caption*{\textbf{Fig. S2.} Relaxation rate measurement of NV Ensembles dispersed on a glass coverslip}
        \label{fig:galaxy2}

\end{figure}

\newpage
\begin{figure}
    \centering
    \includegraphics[width=0.6\textwidth]{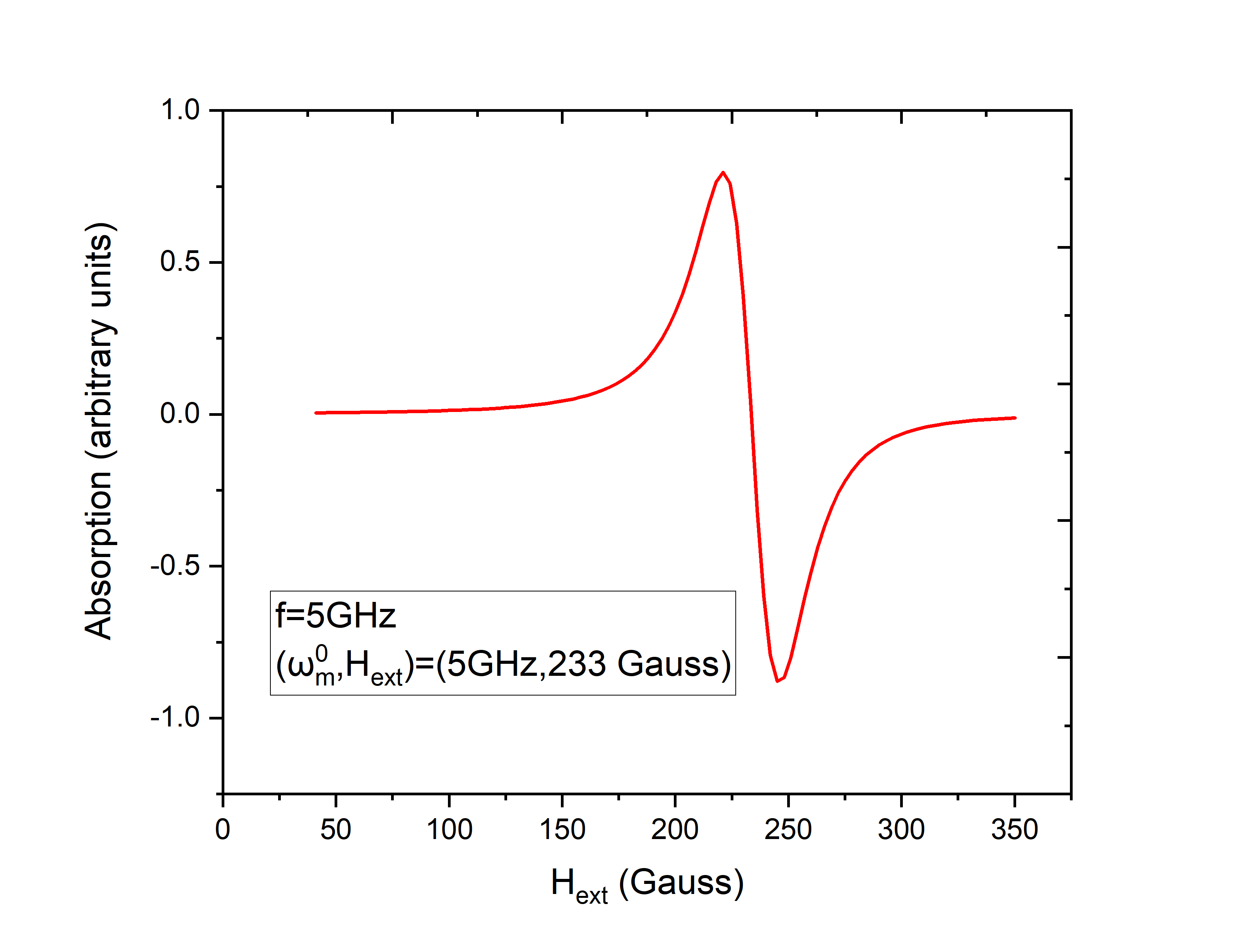}{l}
    \caption*{\textbf{Fig. S3.} Absorption of microwave as a function of magnetic field for a fixed external microwave frequency.}
    \label{fig:galaxy3}
\end{figure}
\newpage
\begin{figure}
    \centering
    \includegraphics[width=0.6\textwidth]{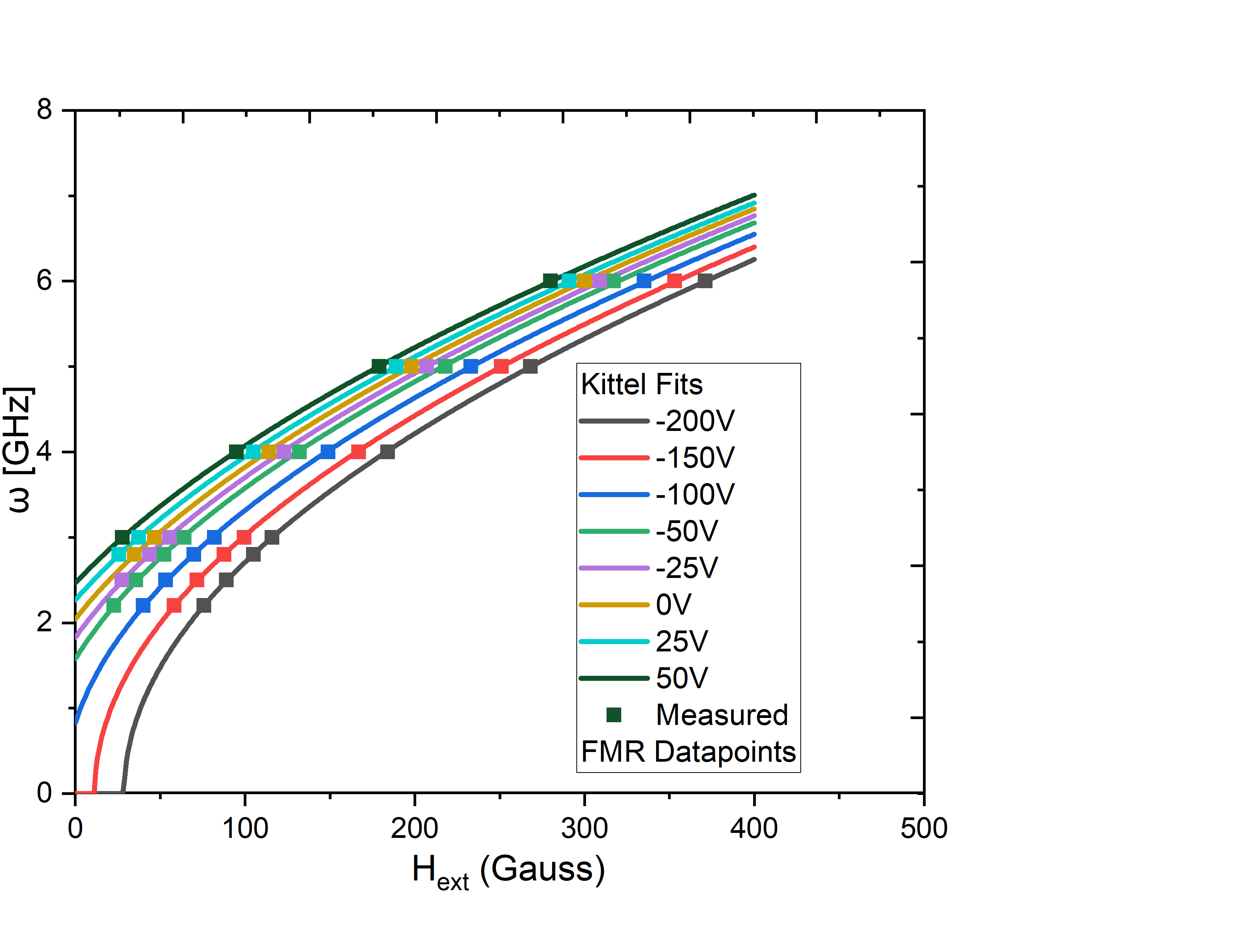}{r}
    \caption*{\textbf{Fig. S4.} FMR Data measured for different voltages from -200V to 50V for $H_{ext}$ applied along $x$-axis. Kittel fits for each dataset are shown as solid lines. The Kittel fit curves are used to extract the uniaxial magnetic anisotropy field $H_{k}$ for different voltages which is presented in inset of Fig. 3(a) in the main text.}
    \label{fig:galaxy4}
\end{figure}

\newpage
\begin{figure}
    \centering
    \includegraphics[width=0.6\textwidth]{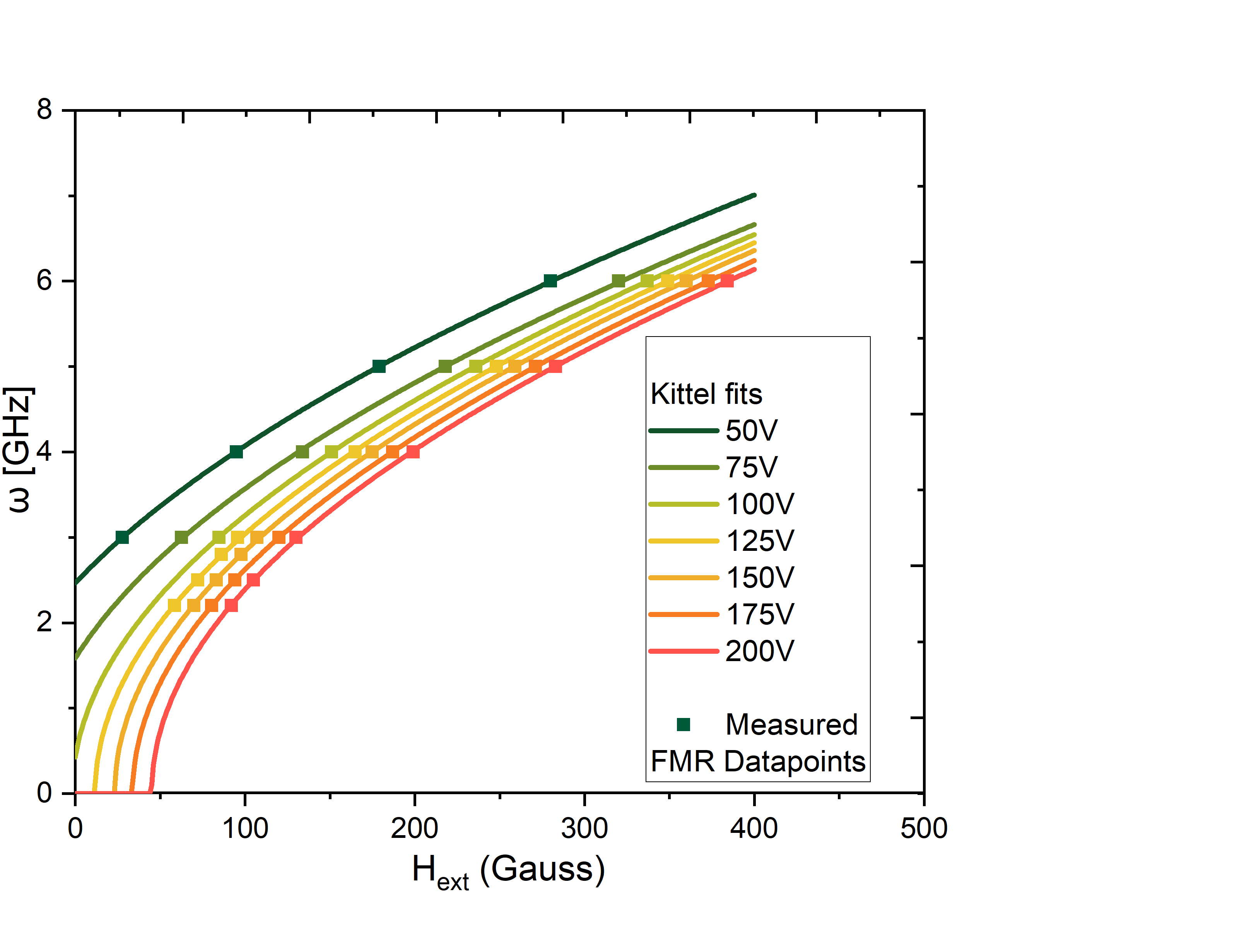}{r}
    \caption*{\textbf{Fig. S5.} FMR Data measured for different voltages from 50V to 200V for $H_{ext}$ applied along $x$-axis. Kittel fits for each dataset are shown as solid lines. The Kittel fit curves are used to extract the uniaxial magnetic anisotropy field $H_{k}$ for different voltages which is presented in inset of Fig. 3(a) in the main text.The Kittel fit curves are also used to extract the FMR frequencies at $H_{ext}$=57Gauss which are presented in Fig. 3(a) in the main text.}
    \label{fig:galaxy5}
\end{figure}
\newpage
\begin{figure}
    \centering
    \includegraphics[width=0.6\textwidth]{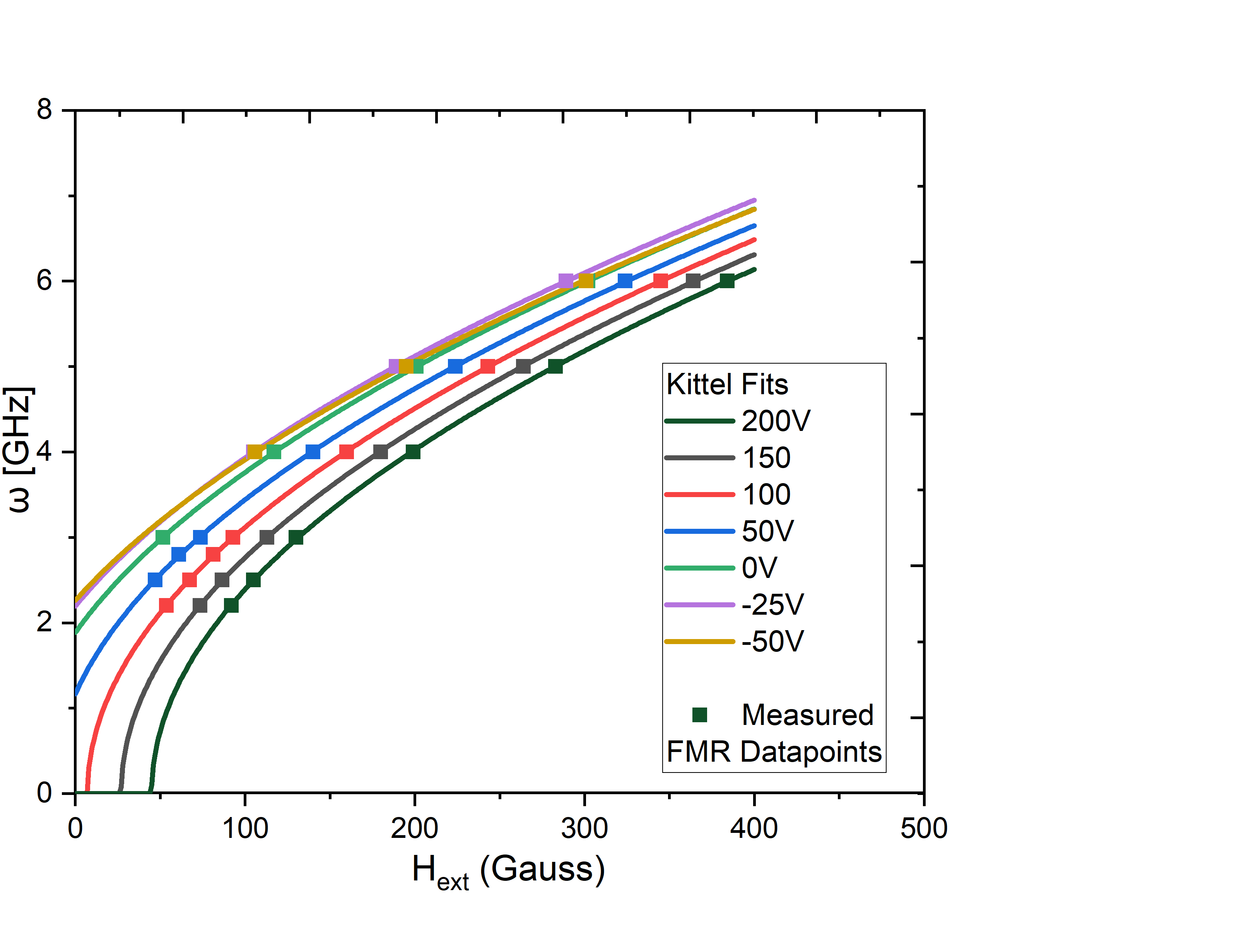}{r}
    \caption*{\textbf{Fig. S6.} FMR Data measured for different voltages from 200V to -50V for $H_{ext}$ applied along $x$-axis. Kittel fits for each dataset are shown as solid lines. The Kittel fit curves are used to extract the uniaxial magnetic anisotropy field $H_{k}$ for different voltages which is presented in inset of Fig. 3(a) in the main text.}
    \label{fig:galaxy6}
\end{figure}
\newpage
\begin{figure}
    \centering
    \includegraphics[width=0.6\textwidth]{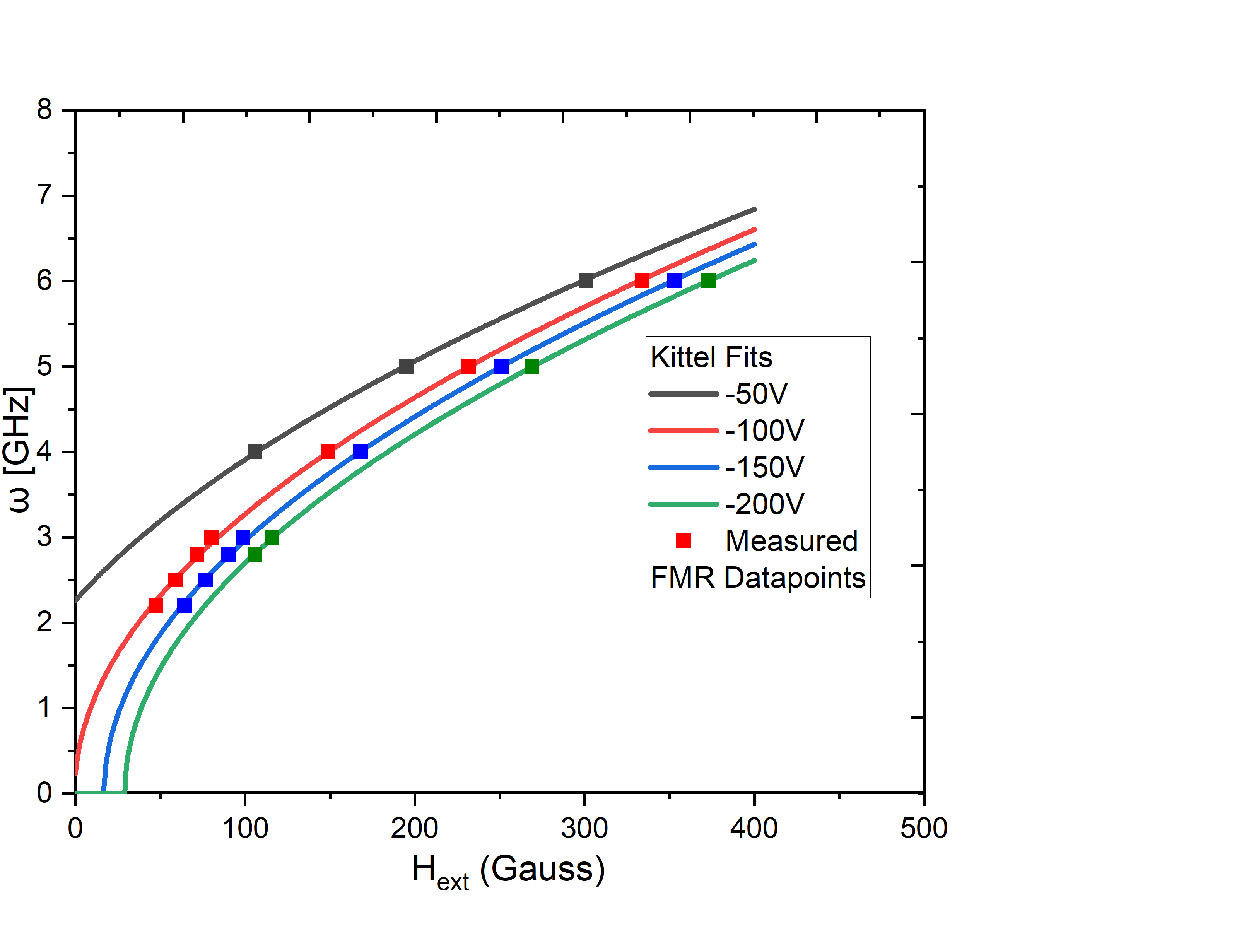}{r}
    \caption*{\textbf{Fig. S7.} FMR Data measured for different voltages from -50V to -200V for $H_{ext}$ applied along $x$-axis. Kittel fits for each dataset are shown as solid lines. The Kittel fit curves are used to extract the uniaxial magnetic anisotropy field $H_{k}$ for different voltages which is presented in inset of Fig. 3(a) in the main text.}
    \label{fig:galaxy7}
\end{figure}

\newpage
\begin{figure}
    \centering
    \includegraphics[width=0.6\textwidth]{PV loop.png}
    \caption*{\textbf{Fig. S8.} Measured normalized electronic polarization ($P_z$) as a function of external DC voltage.}
    \label{fig:galaxy8}
\end{figure}

\newpage
\begin{figure}[h]
    \centering
    \includegraphics[width=0.6\textwidth]{NV on PMN-PT Gamma vs Magnetic field.png}
    \caption*{\textbf{Fig. S9.} Measured relaxation rate of an NV Ensemble on PMN-PT for different magnetic fields applied in the plane of the sample.}
    \label{fig:galaxy9}
\end{figure}

\newpage
\begin{figure}[h]
    \centering
    \includegraphics[width=0.6\textwidth]{NV on PMN-PT Gamma vs Voltage.png}
    \caption*{\textbf{Fig. S10.} Measured relaxation rate of an NV Ensemble on PMN-PT for different voltages.}
    \label{fig:galaxy10}
\end{figure}

\newpage
\begin{figure}[hbtp]
\centering
\includegraphics[width=0.45\textwidth]{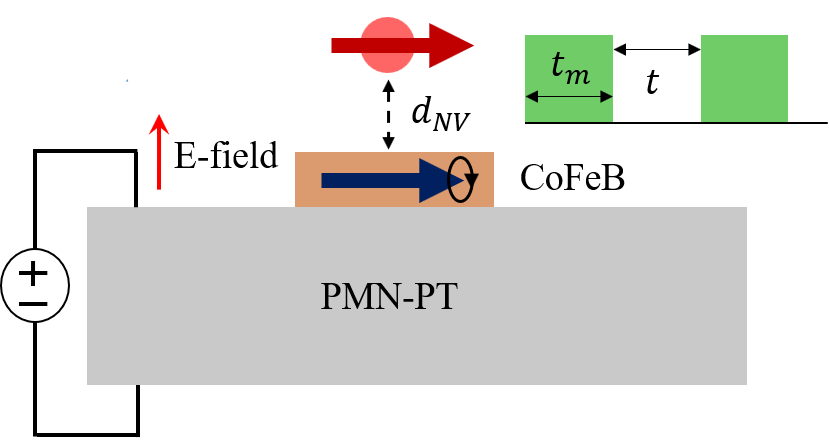}
\caption*{\label{Fig0} \textbf{Fig. S11.} The proposed NV/multiferroic hybrid setup to sense E-field. The schematic shows the pulse sequence.}
\end{figure}

\newpage
\begin{figure}[hbtp] 
\centering
\includegraphics[width=0.45\textwidth]{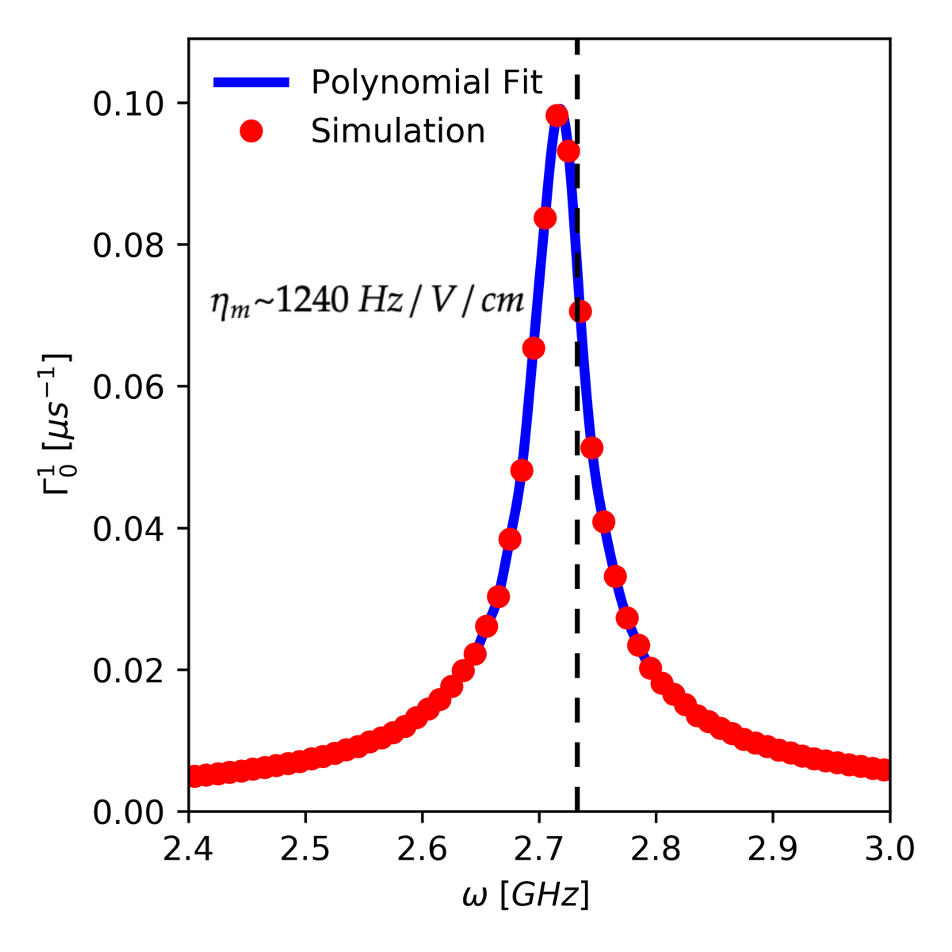}
\caption*{\label{freq_rate} \textbf{Fig. S12.} Numerically calculated NV relaxation rate as a function of FMR frequency which is modulated by external electric field. An external magnetic field of 32 G was applied along the NV axis to bias the NV ESR frequency at the desired value found by maximizing $\eta_m$. The solid line indicates a polynomial fit and the dashed line denotes the bias point. }
\end{figure}